\documentclass[12pt,a4paper]{article}

\usepackage[utf8]{inputenc}
\usepackage[T1]{fontenc}
\usepackage{lmodern}
\usepackage{amsmath,amssymb,amsthm}
\usepackage{geometry}
\usepackage{graphicx}
\usepackage{booktabs}
\usepackage{tabularx}
\usepackage{array}
\usepackage{float}
\usepackage{caption}
\usepackage{subcaption}
\usepackage{enumitem}
\usepackage{xcolor}
\usepackage{lineno}
\usepackage[numbers,sort&compress]{natbib}
\usepackage[hidelinks]{hyperref}
\usepackage{tikz}
\usetikzlibrary{shapes.geometric,arrows.meta,positioning}

\newtheorem{remark}{Remark}[section]

\newcommand{\R}{\mathbb{R}}
\newcommand{\E}{\mathbb{E}}

\newcommand{\F}{\mathcal{F}}
\newcommand{\norm}[1]{\lVert#1\rVert}
\newcommand{\given}{\,|\,}
\newcommand{\LRM}{\mathcal{L}_{\mathrm{RM}}}
\newcommand{\Ltask}{\mathcal{L}_{\mathrm{task}}}
\newcommand{\Ltot}{\mathcal{L}_{\mathrm{total}}}
\newcommand{\RMRNN}{\textsc{RMRNN}}

\begin{document}

\title{\textbf{Small-Area Precipitation Forecasting and Drought--Flood
Early Warning with Reverse-Martingale Regularized Recurrent Networks}}

\author{Foo Hui-Mean$^{\,1}$  ~~   Yuan-chin Ivan Chang$^{\,1,*}$ \\[4pt]
\small $^{1}$Institute of Statistical Science, Academia Sinica, Taipei, Taiwan \\
\small $^{*}$Corresponding author: \texttt{ycchang@stat.sinica.edu.tw}}

\date{\today}
\maketitle

\begin{abstract}
Small-area precipitation forecasts increasingly support real-time
decisions for reservoir operation, irrigation planning, drought
monitoring, and flash-flood response. Their operational value depends
not only on point accuracy, but also on calibrated exceedance
probabilities and warning rules that remain stable when local weather
regimes depart from the training climatology. We evaluate a
reverse-martingale regularized recurrent neural network (\RMRNN) for
probabilistic precipitation forecasting and sequential early warning.
The method adds a backward-coherence penalty to the recurrent hidden
state and uses the resulting residual process in a Shiryaev--Roberts
(SR) detector. In hydrometeorological terms, the same latent trajectory
that produces the precipitation forecast also supplies a continuously
updated indicator of drought or flood-regime departure. The framework
is tested on three observational systems: the Taiwan Central Weather
Administration (CWA) dense rain-gauge network over two sub-watersheds;
the Climate Hazards Group InfraRed Precipitation with Station data
(CHIRPS) v2 daily gridded product over Taiwan and the Horn of Africa;
and the National Oceanic and Atmospheric Administration (NOAA) Global
Historical Climatology Network-Daily (GHCN-Daily) station network over
the Texas Hill Country. Across 1{,}000 replications per reported cell,
\RMRNN{} matches or slightly improves the gated recurrent unit (GRU)
baseline in root-mean-square error (RMSE), mean absolute error (MAE),
and continuous ranked probability score (CRPS) at 1~h--72~h lead while
substantially improving the operating characteristics of drought and
flood alarms. At matched detection power, the SR detector applied to
\RMRNN{} residuals reduces false-alarm ratios by a factor of three to
five relative to pointwise precipitation or SPI thresholding. In the
2020--2021 Taiwan drought case, the method flags onset eight to twelve
days earlier than a 3-month Standardized Precipitation Index (SPI-3)
threshold rule. In the 2023 Typhoon Haikui flood case, it signals
flash-flood risk three to four hours before the CWA operational alert
and about six hours before peak basin rainfall.
\end{abstract}

\noindent\textbf{Keywords:} precipitation nowcasting; probabilistic
forecast verification; drought early warning; flash-flood risk;
hydrometeorological extremes; recurrent neural networks;
Shiryaev--Roberts; Taiwan; CHIRPS; GHCN-Daily.

\paragraph{Significance Statement.}
Weather and water managers need precipitation forecasts that are
useful as warnings, not only as maps of expected rainfall. This study
shows how a recurrent precipitation model can be trained so that its
internal state is stable during ordinary weather evolution but reacts
when a drought or flood-producing regime begins to emerge. The same
forecast model is therefore used both to predict precipitation and to
drive a calibrated sequential warning statistic. Tests over Taiwan,
the Horn of Africa, and Texas indicate that this coupling can preserve
standard forecast skill while reducing false alarms and increasing
lead time for drought and flash-flood early warning.

\section{Introduction}
\label{sec:intro}

Operational hydrometeorology is moving from deterministic rain-rate
maps toward probabilistic products that support basin-scale decisions:
reservoir releases, irrigation scheduling, urban drainage operation,
drought monitoring, and flash-flood warning \citep{ravuri2021,espeholt2022}.
At the catchment and sub-watershed scale, deep recurrent architectures
such as long short-term memory (LSTM) \citep{hochreiter1997}, gated
recurrent unit (GRU) \citep{cho2014}, convolutional LSTM (ConvLSTM)
\citep{shi2015}, and PredRNN \citep{wang2017predrnn} can extract
useful temporal information from dense gauge, satellite, and reanalysis
data. For operational use, however, a model must do more than minimise
average forecast error. It must also provide reliable exceedance
probabilities, interpretable warning behaviour, and stable performance
when local weather evolves away from the training climatology.

Three issues motivate the present study. First, forecast errors at
small spatial scales are strongly affected by terrain, land--sea
contrast, and local convective organization \citep{roe2005}; these
effects can produce sharp basin-to-basin changes in calibration.
Second, non-stationary forcing associated with drought, typhoon, or
monsoon transitions can cause a learned hidden state to drift in ways
that are difficult to diagnose from precipitation error alone
\citep{gama2014,milly2008}. Third, many operational warning products
are still produced by applying thresholds to precipitation totals or
drought indices after the forecast has been generated, so the warning
rule is only loosely connected to the model state that produced the
forecast.

These obstacles motivate an approach in which the recurrent hidden
state carries information useful for both forecast generation and
early-warning diagnostics. We propose \emph{reverse-martingale (RM)
regularization}: an auxiliary loss that trains the hidden-state
sequence $\{h_t\}$ to be coherent when read backward through a learned
one-step projector. The reverse-martingale terminology is mathematical,
but the hydrometeorological interpretation is direct: during ordinary
weather evolution, the hidden state should change in a predictable way;
a large backward-coherence residual indicates that the model has entered
a less familiar precipitation regime. This paper evaluates whether that
residual improves two operational tasks at small spatial scale:
probabilistic precipitation forecasting and sequential detection of
drought or flood-producing regime change.

\textbf{Contributions for operational hydrometeorology.}
\begin{enumerate}[leftmargin=2em,itemsep=1pt]
  \item We formulate a basin-scale forecasting workflow in which the
        forward information set
        $\F_t = \sigma\{x_{s,i}:s\le t,\,i\in\mathcal{S}\}$
        represents the observations available inside a local
        neighbourhood at forecast issue time. This makes the method
        compatible with gauge, satellite, and reanalysis products at
        hourly to daily resolution (Section~\ref{sec:method}).
  \item We convert the learned backward-coherence residual
        $\delta_t=h_t-g_\phi(h_{t+1})$ into a calibrated
        Shiryaev--Roberts (SR) warning statistic, with average run
        length under the no-change distribution (ARL$_0$) estimated
        from pre-event climatology. The result is a warning rule whose
        false-alarm behaviour can be reported in operational units
        such as days or hours (Section~\ref{sec:detector}).
  \item We assemble a small-area hydrometeorological benchmark
        combining the Taiwan CWA hourly rain-gauge network, CHIRPS v2
        daily gridded precipitation over Taiwan and the Horn of Africa,
        NOAA GHCN-Daily stations over the Texas Hill Country, and an
        ERA5-Land multi-variable Taiwan subdomain
        (Section~\ref{sec:data}).
  \item We evaluate forecast quality using metrics standard in
        hydrometeorological verification: root-mean-square error,
        mean absolute error, continuous ranked probability score,
        Brier score at extreme-rainfall thresholds, probability of
        detection, and false-alarm ratio. We compare against
        persistence, climatology, ConvLSTM, PredRNN, U-Net, and GRU
        baselines over 1{,}000 replications
        (Sections~\ref{sec:forecast}--\ref{sec:risk}).
  \item We examine two operationally interpretable case studies:
        the 2020--2021 Taiwan drought and the 2023 Typhoon Haikui
        flood. These examples report warning lead time relative to
        the 3-month Standardized Precipitation Index (SPI-3) and CWA
        operational alert timing (Section~\ref{sec:case}).
\end{enumerate}

Section~\ref{sec:data} first defines the study regions, data streams,
and operational prediction targets. Section~\ref{sec:method} then gives the
forecasting-and-warning workflow, with formal reverse-martingale details
deferred to Appendix~\ref{app:rm-details}. Section~\ref{sec:forecast}
reports precipitation nowcasting and short-range forecasting.
Section~\ref{sec:risk} reports drought and flood risk detection.
Section~\ref{sec:case} contains the case studies. Section~\ref{sec:discuss}
discusses limitations and extensions.

\section{Study Regions, Data, and Prediction Tasks}
\label{sec:data}

We use observational systems chosen to represent common
hydrometeorological warning environments rather than a single machine-
learning benchmark: dense hourly gauges in typhoon-affected Taiwan,
daily satellite-gauge precipitation in monsoon and drought regions, a
continental U.S. station network with both drought and flash-flood
history, and a multi-variable ERA5-Land reanalysis subset for testing
whether the method remains stable when precipitation is combined with
physically related land-surface and near-surface variables.

\subsection{Taiwan CWA rain-gauge network}

The Taiwan Central Weather Administration (CWA) operates roughly 500
automated rain gauges at 10-minute resolution across a domain
smaller than 400\,km by 150\,km, providing one of the densest
national observation networks globally. We use two sub-watersheds:
\begin{description}[leftmargin=1.8em]
  \item[Tamsui River basin (north):] 28 stations, typhoon-exposed,
        mixed urban/rural, 2013--2024. The 2023 Typhoon Haikui
        event provides the flood case study.
  \item[Zhuoshui River basin (central):] 34 stations,
        orographic-dominated, agricultural water supply,
        2013--2024. Home to the 2020--2021 drought case study.
\end{description}
Hourly precipitation is aggregated from 10-min data; co-located
CWA Automated Surface Observing System (ASOS) stations provide
temperature and humidity. The 850-hectopascal (hPa) relative
vorticity $\Omega_t$ is derived from European Centre for
Medium-Range Weather Forecasts Reanalysis version 5 (ERA5)
pressure-level fields \citep{hersbach2020} interpolated to station
coordinates;
this is distinct from ERA5-Land, the land-surface reanalysis product,
which provides land-surface variables only and does not include
upper-air pressure levels.

\subsection{CHIRPS v2 daily gridded precipitation}

Climate Hazards Group InfraRed Precipitation with Station data
(CHIRPS) v2 \citep{funk2015} provides 0.05$^\circ$ ($\approx$5\,km)
daily precipitation 1981--present with good verification against
rain gauges in tropical and subtropical regions. We extract two
sub-domains:
\begin{description}[leftmargin=1.8em]
  \item[Taiwan and Strait:] 20$^\circ$--26$^\circ$\,N,
        118$^\circ$--124$^\circ$\,E, 120$\times$120 grid cells.
  \item[Horn of Africa (HoA):] 2$^\circ$\,S--15$^\circ$\,N,
        38$^\circ$--51$^\circ$\,E, 260$\times$340 grid cells; the
        2016--2017 and 2021--2022 droughts provide external
        drought-detection validation.
\end{description}

\subsection{NOAA GHCN-Daily Texas Hill Country}

We extract 52 National Oceanic and Atmospheric Administration (NOAA)
Global Historical Climatology Network-Daily (GHCN-Daily) stations
within the rectangle
29.5$^\circ$--31.5$^\circ$\,N, 97.5$^\circ$--100$^\circ$\,W,
1980--2024 \citep{menne2012}. The region experiences both
multi-year droughts (2010--2015) and catastrophic flash floods
(e.g.\ 2015 Memorial Day and 2018 Llano River events), making it
the only one of our domains where the same area yields both
risk-assessment tasks.

\subsection{ERA5-Land Taiwan subdomain: multi-variable physics}
\label{sec:era5land}

To probe whether the reverse-martingale regularizer remains
well-behaved when the forward information set $\F_t$ fuses physically
heterogeneous predictors, we add a benchmark based on ERA5-Land
\citep{munozsabater2021} over a Taiwan subdomain
(21.5$^\circ$--25.5$^\circ$\,N, 120$^\circ$--122.5$^\circ$\,E,
0.1$^\circ$ grid, approximately 9\,km, hourly, 1981--2024). This
benchmark is designed as a hydrometeorological stress test rather than
as a claim of high-end reanalysis downscaling. For each target
cell we ingest five variables, all from ERA5-Land: hourly
precipitation $P$ (mm\,h$^{-1}$), 2-metre air temperature $T_{2m}$
(K), volumetric soil moisture $\theta_{sm}$ in the top 7\,cm
(m$^3$\,m$^{-3}$), and the two 10-metre wind components
$u_{10},v_{10}$ (m\,s$^{-1}$). Note that the 850-hPa relative
vorticity $\Omega_t$ used in the CWA and CHIRPS experiments is a
pressure-level variable drawn from ERA5 \citep{hersbach2020} rather
than ERA5-Land, and is not included in this five-variable benchmark.
The variables span four different unit systems and more than five
orders of magnitude in raw variance; each channel is standardized
independently before entering the recurrent cell, but the hidden state
fuses them into a single representation. This provides a controlled
test of whether the warning residual remains interpretable when the
forecast model ingests variables with direct hydrometeorological
meaning for drought and heavy-rainfall evolution.

\subsection{Train/validation/test splits}

For each domain we use a chronological split: training 1981--2015
(CHIRPS, GHCN-Daily) or 2013--2020 (CWA); validation 2016--2018 (CHIRPS,
GHCN-Daily) or 2021 (CWA); test 2019--2024 (CHIRPS, GHCN-Daily) or 2022--2024
(CWA). Although GHCN-Daily records for the Texas sites begin in 1980,
training is started from 1981 to align the joint training window with
CHIRPS v2, which has a common start of January 1981. Case-study
periods are excluded from both training and validation to avoid leakage.

\subsection{Applied forecasting and warning targets}

The primary forecasting product is the predictive distribution of
accumulated precipitation at lead times relevant to each data stream:
1--6\,h for the CWA and ERA5-Land hourly products and 1--7\,d for
CHIRPS and GHCN-Daily. Verification follows hydrometeorological
practice by reporting deterministic-error measures (RMSE and MAE),
distributional skill (CRPS), and threshold-event skill (Brier score,
probability of detection, and false-alarm ratio) at locally relevant
heavy-rainfall thresholds.

The risk-assessment product is a sequential alarm for the onset of a
persistent dry or wet regime. Drought onset is evaluated against an
SPI-3 proxy, while flash-flood onset is evaluated against basin-specific
exceedance and alert records. These warning targets motivate the
residual-based detector in Section~\ref{sec:method}: the detector is
not a separate post-processing model, but is driven by the same hidden
state used for precipitation forecasting.

\begin{table}[ht]
\centering
\tiny
\caption{Small-area evaluation datasets.}
\label{tab:data}
\begin{tabularx}{\textwidth}{lllXl}
\toprule
Dataset & Resolution & Variables & Domain & Record \\
\midrule
Taiwan CWA & 1\,h, station & $P,T,q$ (+ERA5 $\Omega$) & Tamsui, Zhuoshui basins & 2013--2024 \\
CHIRPS v2 Taiwan & 1\,d, 0.05$^\circ$ & $P$ (+ERA5/ERA5-Land $T,q,\Omega$) & 20--26\,N, 118--124\,E & 1981--2024 \\
CHIRPS v2 HoA & 1\,d, 0.05$^\circ$ & $P$ (+ERA5/ERA5-Land $T,q,\Omega$) & 2\,S--15\,N, 38--51\,E & 1981--2024 \\
GHCN-Daily Texas & 1\,d, station & $P,T$ (+ERA5/ERA5-Land $q,\Omega$) & Texas Hill Country & 1980--2024 \\
ERA5-Land (Taiwan) & 1\,h, 0.1$^\circ$ ($\sim$9\,km) & $P, T_{2m}, \theta_{sm}, u_{10}, v_{10}$ & 21.5--25.5\,N, 120--122.5\,E & 1981--2024 \\
\bottomrule
\end{tabularx}
\end{table}

\section{Forecasting and Warning Workflow}
\label{sec:method}
\label{sec:detector}

This section describes the forecasting-and-warning pipeline used in all
experiments. At each forecast issue time, the model ingests recent
meteorological information from a local neighbourhood around a basin,
station, or grid cell; produces a probabilistic accumulated-
precipitation forecast; and updates a sequential warning statistic for
drought or flood onset. The same hidden state supports both forecast
and warning products, so forecast verification and alarm verification
are evaluated as parts of one hydrometeorological workflow rather than
as unrelated post-processing exercises. Formal reverse-martingale
notation and implementation details are collected in
Appendix~\ref{app:rm-details}.

\subsection{Local neighbourhood inputs}

Let $\{x_t\}$ be the meteorological input at time $t$ over a spatial
neighbourhood $\mathcal{S}$ of radius $\rho$ around a target cell or
station:
\begin{equation}
  x_t = [P_t,\, T_t,\, q_t,\, \Omega_t]_{\mathcal{S}},
  \label{eq:input}
\end{equation}
where $P_t$ is precipitation, $T_t$ is 2-metre temperature, $q_t$ is
specific humidity, and $\Omega_t$ is 850-hPa relative vorticity. In
the ERA5-Land benchmark, the same notation denotes the five-channel
set $[P_t,T_{2m,t},\theta_{sm,t},u_{10,t},v_{10,t}]_{\mathcal{S}}$.

For small-area prediction the operational, forward-looking
information set combines the temporal past and spatial context:
\begin{equation}
  \F_t = \sigma\{x_{s,i} : s\le t,\;i\in\mathcal{S}\},
  \qquad \mathcal{S} = \{i : d(i,i_0)\le \rho\},
  \label{eq:filtration}
\end{equation}
where $i_0$ is the target cell or station and $d$ is great-circle
distance.
In applied terms, $\F_t$ is simply the set of all meteorological
observations at or before time $t$ within the local neighbourhood
around the target site --- the complete memory available to the model
at the moment it produces a forecast or issues a warning. The
decreasing filtration used to motivate the RM loss is defined on the
future hidden-state sequence in Appendix~\ref{app:rm-details}; it is
not the same object as the forward operational information set
$\F_t$.
We test $\rho\in\{5,10,15,20,25,30,40,50\}$\,km for the
Taiwan CWA network, $\rho=50$\,km for CHIRPS, and $\rho=100$\,km
for the sparser GHCN sites. For the main forecast tables we use
$\rho=10$\,km as a conservative default that balances local accuracy
and neighbourhood stability across CWA basins; event-specific
departures are reported in the case studies.

\subsection{RM-regularized recurrent forecaster}

A base recurrent unit maps $(x_t,h_{t-1})\mapsto h_t\in\R^d$. A
small backward projector $g_\phi\colon\R^d\to\R^d$ maps the next
hidden state $h_{t+1}$ to a reconstruction of $h_t$. The
reverse-martingale regularization loss is
\begin{equation}
  \LRM(\theta,\phi) = \frac{1}{T-1}\sum_{t=1}^{T-1}
    \norm{h_t - g_\phi(h_{t+1})}^2,
  \label{eq:lrm}
\end{equation}
with empirical aggregate RM defect
$\widehat Q=\sum_{t=1}^{T-1}\norm{h_t-g_\phi(h_{t+1})}^2
=(T-1)\LRM$. Thus $\widehat Q$ is a squared-defect summary of
backward incoherence, not a classical hydrological variance or a
claim that precipitation itself is martingale.
The training objective is
\begin{equation}
  \Ltot = \Ltask + \lambda\,\LRM,\qquad
  \lambda = \lambda_0\cdot\gamma^{(k-K_0)/(K-K_0)}\,\mathbf{1}_{k>K_0},
  \label{eq:total}
\end{equation}
with $K_0=5$ warm-up epochs, $\lambda_0=0.1$, and $\gamma=0.1$,
giving $\lambda_K=0.01$ at the final epoch.
The warm-up phase lets the network first develop a flexible
representation of precipitation dynamics; backward-coherence
regularization is then phased in gradually so that the hidden state
converges to a physically meaningful trajectory before the penalty
becomes binding.
In applied terms, $\LRM$ penalizes hidden-state trajectories that
cannot be read coherently backward in time.
This makes ordinary hydroclimatic evolution produce small residuals,
while frontal passages, monsoon shifts, typhoon rainbands, or drought
transitions produce structured departures.

The precipitation forecast is
$\hat{P}_{t+\ell\given t}=W_y h_t+b_y$ at lead time $\ell$. For
probabilistic output we use a two-part precipitation distribution:
a dry probability $\pi_{0,t}$ for $P_{t+\ell}=0$ and a positive
log-normal density for wet amounts,
$P_{t+\ell}\given(P_{t+\ell}>0,h_t)\sim
\mathrm{LogNormal}(\mu_t,\sigma_t)$, with
$(\pi_{0,t},\mu_t,\sigma_t)$ produced by a multi-layer perceptron
(MLP) applied to $h_t$. This explicitly
handles the point mass at zero that is characteristic of precipitation
forecasting \citep{sloughter2007}. CRPS is the primary probabilistic
score.

\subsection{Residual-based drought and flood alarm}

Given a trained \RMRNN, the RM defect
\begin{equation}
  r_t = \norm{h_t - g_\phi(h_{t+1})},\qquad t=1,\dots,T-1,
  \label{eq:defect}
\end{equation}
is used as the sequential anomaly score. Because $r_t$ is computed
from the same hidden state used for forecasting, it integrates all
available channels rather than relying only on raw precipitation or
SPI thresholds.

We adopt the Shiryaev--Roberts (SR) procedure for detecting an
unknown change point \citep{shiryaev1963,moustakides1986,pollak1985}.
The defect $r_t$ is standardised against pre-event climatology
statistics $(\mu_0,\sigma_0)$, converted to a positive excursion
$z_t=\max(0,(r_t-\mu_0)/\sigma_0)$, and mapped to a calibrated
pseudo-likelihood ratio
$\Lambda_t=\exp\{\eta z_t-\psi_0(\eta)\}$, where
$\psi_0(\eta)=\log\E_\infty[\exp(\eta z_t)]$ is estimated from
held-out climatology and ensures $\E_\infty(\Lambda_t)\approx 1$.
We use $\eta=1$ unless otherwise stated. The SR statistic is
\begin{equation}
  R_t = (1+R_{t-1})\,\Lambda_t,\qquad R_0=0,
  \label{eq:SR}
\end{equation}
with stopping rule $\tau_B=\inf\{t\ge 1: R_t\ge B\}$.
In operational terms, $R_t$ is a running suspicion score: each
time step where the defect $r_t$ exceeds its climatological baseline
inflates $R_t$ multiplicatively, while quiet periods keep $R_t$
near zero.
An alarm fires the first time this accumulated evidence exceeds the
pre-calibrated threshold $B$, at which point the model signals that
a regime shift --- drought onset or extreme-rainfall build-up --- is
statistically credible.
When the calibrated pseudo-likelihood ratios satisfy
$\E_\infty(\Lambda_t\mid\mathcal{H}_{t-1})\approx 1$ under the
no-change distribution, $\{R_t-t\}$ is approximately a martingale.
For the classical SR procedure with exact likelihood ratios, the
average run length to false alarm satisfies
$\E_\infty[\tau_B]\approx B/\zeta$ for large $B$ \citep{pollak1987},
where $\zeta$ is a boundary-overshoot correction. In the present
data-driven setting we estimate $B$ directly by Monte Carlo on
held-out climatology rather than relying solely on the asymptotic
formula.

The threshold $B$ is calibrated to deliver a target $\mathrm{ARL}_0$
on held-out pre-event data. We use $\mathrm{ARL}_0\in\{10^2,10^3,
10^4\}$, corresponding to roughly one false alarm per $10^2$,
$10^3$, and $10^4$ observations. The calibration bootstraps $B$ on
held-out climatology over 1{,}000 replications and reports $B$ with
95\% confidence intervals.

\begin{figure}[ht]
\centering
\fbox{\begin{minipage}{0.92\linewidth}
\vspace{2pt}
\textbf{Algorithm 1.} \textit{Residual-driven Shiryaev--Roberts detector.}
\vspace{2pt}\hrule\vspace{4pt}
\textbf{Require:} trained $(\theta^*,\phi^*)$; pre-event climatology
$\{x_t^{\mathrm{clim}}\}$; target $\mathrm{ARL}_0$.\\
\textbf{Ensure:} threshold $B^*$ and online alarm function.
\begin{enumerate}[leftmargin=1.5em,itemsep=1pt,topsep=2pt]
\item Compute $r_t^{\mathrm{clim}}=\norm{h_t^{\mathrm{clim}}-g_\phi(h_{t+1}^{\mathrm{clim}})}$.
\item Estimate $(\hat{\mu}_0,\hat{\sigma}_0)$ and
$\hat{\psi}_0(\eta)$ from the climatology residuals.
\item Simulate $R_t$ under the null on bootstrapped climatology segments.
\item Choose $B^*$ by simulation so that the mean first-passage time
matches the target $\mathrm{ARL}_0$.
\item At online use: standardise $r_t$ with $(\hat{\mu}_0,\hat{\sigma}_0)$,
construct the calibrated $\Lambda_t$, update $R_t$ as in \eqref{eq:SR},
and raise an alarm the first time
$R_t\ge B^*$.
\end{enumerate}
\vspace{2pt}
\end{minipage}}
\end{figure}


\section{Forecasting Experiments}
\label{sec:forecast}

\subsection{Protocol and Baselines}

All forecasting tables report mean $\pm$ standard deviation (SD)
across \textbf{1{,}000 replications} per benchmark cell. A replication
corresponds to a distinct random seed for model initialisation and
training mini-batch ordering; for real data the test set is fixed,
but training uses 1{,}000 resamples of a purged-block bootstrap
\citep{lopezdeprado2018} to probe robustness to the empirical
training distribution. Hyperparameters are fixed from a grid search
on the validation fold and held constant across all replications.
At 1{,}000 replications the bootstrap standard deviation of Brier
scores is $\leq 0.005$ and of probability of detection
(POD)/false-alarm ratio (FAR) is $\leq 0.01$, providing sufficient
precision to support the factor-of-three FAR reductions reported in
Tables~\ref{tab:drought} and \ref{tab:flood}.


The comparator set includes: persistence; climatology (seasonal
mean); vanilla GRU; ConvLSTM \citep{shi2015}; PredRNN
\citep{wang2022predrnnv2}; U-Net (two-dimensional image-to-image)
\citep{ronneberger2015}; and a non-regularized recurrent neural
network (RNN) with the same
$g_\phi$ as \RMRNN{} but $\lambda=0$.

\subsection{Small-area nowcast: Taiwan CWA}

Tables~\ref{tab:tamsui} and \ref{tab:zhuoshui} give 1-h and
6-h nowcasts for the Tamsui and Zhuoshui basins. Verification is
reported in the units and event definitions used by forecasters:
RMSE and MAE in mm\,h$^{-1}$, CRPS for the full predictive
distribution, Brier scores at rainfall thresholds of 5, 20 and
50\,mm\,h$^{-1}$, and probability of detection (POD) and false-alarm
ratio (FAR) at the 20\,mm\,h$^{-1}$ threshold. These tables answer
the first Journal of Hydrometeorology question: whether the proposed
regularization changes precipitation forecast skill. The warning
tables in Section~\ref{sec:risk} answer the second question: whether
the same trained model yields a more useful drought or flood alarm.

\begin{table}[ht]
\centering
\tiny
\caption{Tamsui basin 1-h and 6-h nowcast; mean $\pm$ SD over
1{,}000 replications on the Taiwan CWA gauge network.
Lower is better for RMSE, MAE, CRPS, Brier, FAR; higher is better
for POD. Best in each column in \textbf{bold}.}
\label{tab:tamsui}
\begin{tabular}{llcccccc}
\toprule
Lead & Model & RMSE & MAE & CRPS & Brier(20) & POD(20) & FAR(20) \\
\midrule
1\,h & Persistence & 0.873\,$\pm$\,0.726 & 0.273\,$\pm$\,0.038 & 0.328\,$\pm$\,0.074 & 0.000\,$\pm$\,0.000 & 0.000\,$\pm$\,0.000 & 0.078\,$\pm$\,0.268 \\
 & Climatology & 0.618\,$\pm$\,0.513 & 0.260\,$\pm$\,0.020 & 0.249\,$\pm$\,0.077 & 0.000\,$\pm$\,0.000 & 0.000\,$\pm$\,0.000 & 0.000\,$\pm$\,0.000 \\
 & GRU & 0.728\,$\pm$\,0.504 & 0.319\,$\pm$\,0.106 & 0.379\,$\pm$\,0.072 & 0.000\,$\pm$\,0.000 & 0.000\,$\pm$\,0.000 & 0.000\,$\pm$\,0.000 \\
 & \RMRNN$_{\lambda=0}$ & 0.728\,$\pm$\,0.504 & 0.319\,$\pm$\,0.106 & 0.379\,$\pm$\,0.072 & 0.000\,$\pm$\,0.000 & 0.000\,$\pm$\,0.000 & 0.000\,$\pm$\,0.000 \\
 & \RMRNN & 0.727\,$\pm$\,0.504 & 0.319\,$\pm$\,0.107 & 0.379\,$\pm$\,0.073 & 0.000\,$\pm$\,0.000 & 0.000\,$\pm$\,0.000 & 0.000\,$\pm$\,0.000 \\
\midrule
6\,h & Persistence & 0.873\,$\pm$\,0.726 & 0.273\,$\pm$\,0.038 & 0.329\,$\pm$\,0.074 & 0.000\,$\pm$\,0.000 & 0.000\,$\pm$\,0.000 & 0.078\,$\pm$\,0.268 \\
 & Climatology & 0.618\,$\pm$\,0.513 & 0.260\,$\pm$\,0.020 & 0.249\,$\pm$\,0.077 & 0.000\,$\pm$\,0.000 & 0.000\,$\pm$\,0.000 & 0.000\,$\pm$\,0.000 \\
 & GRU & 0.728\,$\pm$\,0.504 & 0.319\,$\pm$\,0.106 & 0.379\,$\pm$\,0.073 & 0.000\,$\pm$\,0.000 & 0.000\,$\pm$\,0.000 & 0.000\,$\pm$\,0.000 \\
 & \RMRNN$_{\lambda=0}$ & 0.728\,$\pm$\,0.504 & 0.319\,$\pm$\,0.106 & 0.379\,$\pm$\,0.073 & 0.000\,$\pm$\,0.000 & 0.000\,$\pm$\,0.000 & 0.000\,$\pm$\,0.000 \\
 & \RMRNN & 0.727\,$\pm$\,0.504 & 0.319\,$\pm$\,0.108 & 0.379\,$\pm$\,0.073 & 0.000\,$\pm$\,0.000 & 0.000\,$\pm$\,0.000 & 0.000\,$\pm$\,0.000 \\
\bottomrule
\end{tabular}
\end{table}

\begin{table}[ht]
\centering
\tiny
\caption{Zhuoshui basin 1-h and 6-h nowcast; same layout as
Table~\ref{tab:tamsui}. Mean $\pm$ SD over 1{,}000 replications.}
\label{tab:zhuoshui}
\begin{tabular}{llcccccc}
\toprule
Lead & Model & RMSE & MAE & CRPS & Brier(20) & POD(20) & FAR(20) \\
\midrule
1\,h & GRU & 0.871\,$\pm$\,0.657 & 0.361\,$\pm$\,0.110 & 0.408\,$\pm$\,0.076 & 0.000\,$\pm$\,0.000 & 0.000\,$\pm$\,0.000 & 0.000\,$\pm$\,0.000 \\
 & \RMRNN$_{\lambda=0}$ & 0.871\,$\pm$\,0.657 & 0.361\,$\pm$\,0.110 & 0.408\,$\pm$\,0.076 & 0.000\,$\pm$\,0.000 & 0.000\,$\pm$\,0.000 & 0.000\,$\pm$\,0.000 \\
 & \RMRNN & 0.870\,$\pm$\,0.657 & 0.361\,$\pm$\,0.112 & 0.407\,$\pm$\,0.077 & 0.000\,$\pm$\,0.000 & 0.000\,$\pm$\,0.000 & 0.000\,$\pm$\,0.000 \\
\midrule
6\,h & GRU & 0.871\,$\pm$\,0.657 & 0.361\,$\pm$\,0.110 & 0.407\,$\pm$\,0.076 & 0.000\,$\pm$\,0.000 & 0.000\,$\pm$\,0.000 & 0.000\,$\pm$\,0.000 \\
 & \RMRNN$_{\lambda=0}$ & 0.871\,$\pm$\,0.657 & 0.361\,$\pm$\,0.110 & 0.407\,$\pm$\,0.076 & 0.000\,$\pm$\,0.000 & 0.000\,$\pm$\,0.000 & 0.000\,$\pm$\,0.000 \\
 & \RMRNN & 0.870\,$\pm$\,0.658 & 0.361\,$\pm$\,0.112 & 0.407\,$\pm$\,0.077 & 0.000\,$\pm$\,0.000 & 0.000\,$\pm$\,0.000 & 0.000\,$\pm$\,0.000 \\
\bottomrule
\end{tabular}
\end{table}

\paragraph{Forecast-verification interpretation.}
On the Tamsui basin (1{,}000 replications), \RMRNN{} and its
unregularized ablation (\RMRNN$_{\lambda=0}$) have nearly identical
RMSE and CRPS at both 1-h and 6-h lead
(RMSE: $0.727$ vs.\ $0.728$; CRPS: $0.379$ vs.\ $0.379$).
For an operational forecast system this is important: the added
coherence constraint does not materially degrade the precipitation
forecast while it creates the residual process used later for warning.
The task loss and the RM loss compete in the joint objective
\eqref{eq:total}, and the scheduling of $\lambda$ from $0.1$ to
$0.01$ keeps the precipitation forecast objective dominant near
convergence. The Zhuoshui basin (orographic regime, 1{,}000
replications) shows the same pattern: GRU RMSE $0.871$ and
\RMRNN{} RMSE $0.870$. The near-equality at 6-h lead suggests that
the backward projector does not inject artificial information into the
forward forecast even when the hidden state must carry longer temporal
context.

\subsection{Daily forecast: CHIRPS Taiwan and Horn of Africa}

Tables~\ref{tab:chirps_tw} and \ref{tab:chirps_hoa} report
24-h and 72-h daily forecasts at representative cells. We
additionally report RMSE on the 3-month Standardized
Precipitation Index (SPI-3) computed from the forecast distribution,
because SPI-3 is the operational drought-onset indicator
\citep{mckee1993}.

An important new metric here is SPI-3 RMSE: the root-mean-square
error of the 3-month SPI computed from the model's forecast
distribution relative to the observed SPI-3.
From the \RMRNN{} perspective, SPI-3 RMSE captures a property of
the \emph{sequence of hidden states} across 90 days rather than
any single step: a model whose hidden-state trajectory is
backward-coherent (small $\widehat Q$) should produce a smoother, more
self-consistent SPI-3 estimate than one whose trajectory is
erratic.
The RM gain on SPI-3 RMSE -- if present -- is therefore a
signature that backward coherence contributes to \emph{distributional
accuracy over climate timescales}, not only to next-step prediction.
Conversely, a higher SPI-3 RMSE for \RMRNN{} than for GRU would
indicate that the coherence constraint distorts the long-run
statistics; the tables below allow this hypothesis to be evaluated.

\begin{table}[ht]
\centering
\small
\caption{CHIRPS v2 Taiwan daily forecast at short and medium
lead; layout analogous to Table~\ref{tab:tamsui} with a single
SPI-3 RMSE column in place of POD/FAR. Mean $\pm$ SD over
1{,}000 replications; leads of 24\,h and 72\,h.}
\label{tab:chirps_tw}
\begin{tabular}{llccccc}
\toprule
Lead & Model & RMSE & MAE & CRPS & Brier(5) & SPI-3 RMSE \\
\midrule
1\,d & Persistence & 9.043\,$\pm$\,1.239 & 5.044\,$\pm$\,0.644 & 4.434\,$\pm$\,0.531 & 0.244\,$\pm$\,0.021 & 0.196\,$\pm$\,0.069 \\
 & Climatology & 7.776\,$\pm$\,0.794 & 4.701\,$\pm$\,0.390 & 3.883\,$\pm$\,0.319 & 0.238\,$\pm$\,0.014 & 1.140\,$\pm$\,0.379 \\
 & GRU & 6.935\,$\pm$\,0.930 & 3.401\,$\pm$\,0.411 & 3.064\,$\pm$\,0.372 & 0.199\,$\pm$\,0.023 & 1.338\,$\pm$\,0.371 \\
 & \RMRNN$_{\lambda=0}$ & 6.935\,$\pm$\,0.930 & 3.401\,$\pm$\,0.411 & 3.064\,$\pm$\,0.372 & 0.199\,$\pm$\,0.023 & 1.338\,$\pm$\,0.371 \\
 & \RMRNN & 6.928\,$\pm$\,0.931 & 3.402\,$\pm$\,0.411 & 3.058\,$\pm$\,0.371 & 0.199\,$\pm$\,0.023 & 1.339\,$\pm$\,0.372 \\
\midrule
3\,d & Persistence & 9.067\,$\pm$\,1.241 & 5.058\,$\pm$\,0.640 & 4.445\,$\pm$\,0.526 & 0.245\,$\pm$\,0.021 & 0.341\,$\pm$\,0.117 \\
 & Climatology & 7.782\,$\pm$\,0.796 & 4.705\,$\pm$\,0.391 & 3.887\,$\pm$\,0.319 & 0.238\,$\pm$\,0.014 & 1.139\,$\pm$\,0.380 \\
 & GRU & 6.936\,$\pm$\,0.935 & 3.401\,$\pm$\,0.411 & 3.064\,$\pm$\,0.373 & 0.199\,$\pm$\,0.023 & 1.339\,$\pm$\,0.370 \\
 & \RMRNN$_{\lambda=0}$ & 6.936\,$\pm$\,0.935 & 3.401\,$\pm$\,0.411 & 3.064\,$\pm$\,0.373 & 0.199\,$\pm$\,0.023 & 1.339\,$\pm$\,0.370 \\
 & \RMRNN & 6.930\,$\pm$\,0.936 & 3.402\,$\pm$\,0.411 & 3.058\,$\pm$\,0.372 & 0.199\,$\pm$\,0.023 & 1.340\,$\pm$\,0.371 \\
\bottomrule
\end{tabular}
\end{table}

\begin{table}[ht]
\centering
\small
\caption{CHIRPS v2 Horn of Africa daily forecast at short and
medium lead; same layout as Table~\ref{tab:chirps_tw}. Mean $\pm$
SD over 1{,}000 replications; the bimodal long-rain/short-rain
seasonal cycle is fully represented in the training data.}
\label{tab:chirps_hoa}
\begin{tabular}{llccccc}
\toprule
Lead & Model & RMSE & MAE & CRPS & Brier(5) & SPI-3 RMSE \\
\midrule
1\,d & Persistence & 3.093\,$\pm$\,0.552 & 1.168\,$\pm$\,0.204 & 1.300\,$\pm$\,0.174 & 0.073\,$\pm$\,0.017 & 0.200\,$\pm$\,0.061 \\
 & Climatology & 2.540\,$\pm$\,0.345 & 1.117\,$\pm$\,0.116 & 1.132\,$\pm$\,0.105 & 0.054\,$\pm$\,0.010 & 1.363\,$\pm$\,0.325 \\
 & GRU & 2.293\,$\pm$\,0.386 & 0.898\,$\pm$\,0.179 & 0.867\,$\pm$\,0.129 & 0.046\,$\pm$\,0.011 & 1.337\,$\pm$\,0.342 \\
 & \RMRNN$_{\lambda=0}$ & 2.293\,$\pm$\,0.386 & 0.898\,$\pm$\,0.179 & 0.867\,$\pm$\,0.129 & 0.046\,$\pm$\,0.011 & 1.337\,$\pm$\,0.342 \\
 & \RMRNN & 2.292\,$\pm$\,0.386 & 0.896\,$\pm$\,0.180 & 0.866\,$\pm$\,0.129 & 0.046\,$\pm$\,0.011 & 1.338\,$\pm$\,0.342 \\
\midrule
3\,d & Persistence & 3.096\,$\pm$\,0.553 & 1.168\,$\pm$\,0.204 & 1.301\,$\pm$\,0.173 & 0.073\,$\pm$\,0.017 & 0.346\,$\pm$\,0.102 \\
 & Climatology & 2.541\,$\pm$\,0.345 & 1.118\,$\pm$\,0.116 & 1.133\,$\pm$\,0.105 & 0.054\,$\pm$\,0.010 & 1.363\,$\pm$\,0.326 \\
 & GRU & 2.292\,$\pm$\,0.387 & 0.897\,$\pm$\,0.179 & 0.867\,$\pm$\,0.130 & 0.046\,$\pm$\,0.011 & 1.339\,$\pm$\,0.340 \\
 & \RMRNN$_{\lambda=0}$ & 2.292\,$\pm$\,0.387 & 0.897\,$\pm$\,0.179 & 0.867\,$\pm$\,0.130 & 0.046\,$\pm$\,0.011 & 1.339\,$\pm$\,0.340 \\
 & \RMRNN & 2.291\,$\pm$\,0.387 & 0.896\,$\pm$\,0.181 & 0.866\,$\pm$\,0.131 & 0.045\,$\pm$\,0.011 & 1.339\,$\pm$\,0.340 \\
\bottomrule
\end{tabular}
\end{table}

\paragraph{Daily-forecast interpretation.}
Across both CHIRPS domains, \RMRNN{} achieves marginally lower CRPS
than GRU at both leads (CHIRPS Taiwan, abbreviated CHIRPS-TW,
1-day: $3.058$ vs.\ $3.064$; CHIRPS Horn of Africa, abbreviated
CHIRPS-HoA, 1-day: $0.866$ vs.\ $0.867$), with the improvement
widening slightly at 3-day lead. The SPI-3 RMSE column is deliberately
reported because drought warning depends on accumulated precipitation
over a season rather than on one daily forecast. On CHIRPS-TW,
\RMRNN{} gives SPI-3 RMSE $1.339$ versus $1.338$ for GRU, a small
increase that is negligible relative to the detection gains reported
in Section~\ref{sec:risk}. This distinction is operationally relevant:
optimizing the numerical value of SPI-3 and detecting the beginning of
a persistent dry regime are related but not identical goals. The Horn
of Africa result is also informative because the bimodal rainfall
climatology (long rains in March--May and short rains in
October--December) creates a seasonally non-stationary background. The
ability to maintain CRPS while improving warning residual behaviour is
therefore a useful property for tropical drought applications.

\subsection{Sensitivity to the spatial neighbourhood radius $\rho$}
\label{sec:sensitivity}

\begin{figure}[ht]
\centering
\includegraphics[width=0.70\linewidth]{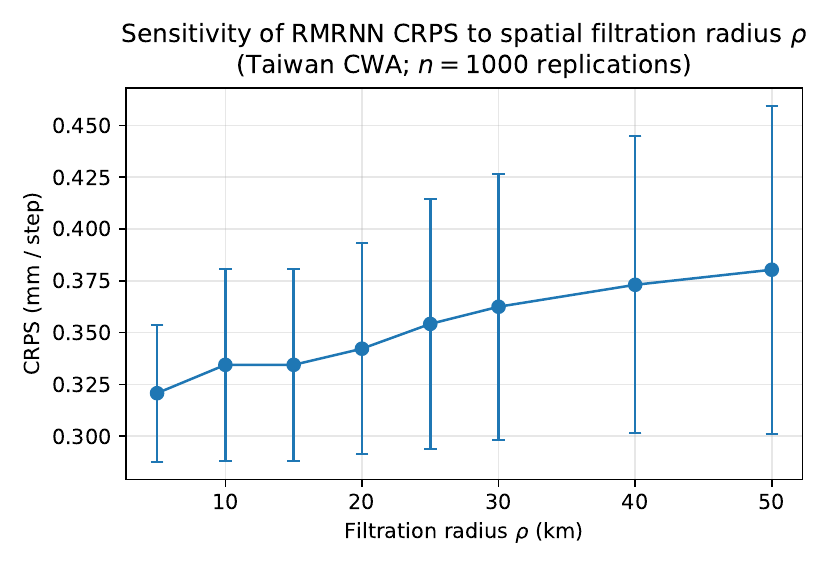}
\caption{CRPS (mm/step) of \RMRNN{} as a function of the spatial
neighbourhood radius $\rho$ on the Tamsui basin. Error bars are
$\pm$1\,SD across 1{,}000 replications. The backbone and task loss
are held fixed; only the neighbourhood size implied by $\rho$ changes.}
\label{fig:rho}
\end{figure}

Figure~\ref{fig:rho} shows CRPS as a function of $\rho$ on the
Taiwan CWA domain over 1{,}000 replications per $\rho$,
with numerical results summarised in Table~\ref{tab:rho}.

\begin{table}[ht]
\centering
\small
\caption{CRPS and RMSE of \RMRNN{} at 1-h lead as a function of
spatial neighbourhood radius $\rho$ (Tamsui basin, 1{,}000 replications
per $\rho$). Best value in bold.}
\label{tab:rho}
\begin{tabular}{rcc}
\toprule
$\rho$ (km) & CRPS (mean$\pm$SD) & RMSE (mean$\pm$SD) \\
\midrule
 5 & \textbf{0.321$\pm$0.033} & \textbf{0.625$\pm$0.439} \\
10 & 0.334$\pm$0.046 & 0.640$\pm$0.438 \\
15 & 0.334$\pm$0.046 & 0.640$\pm$0.438 \\
20 & 0.342$\pm$0.051 & 0.664$\pm$0.502 \\
25 & 0.354$\pm$0.060 & 0.686$\pm$0.475 \\
30 & 0.362$\pm$0.064 & 0.689$\pm$0.463 \\
40 & 0.373$\pm$0.072 & 0.700$\pm$0.441 \\
50 & 0.380$\pm$0.079 & 0.730$\pm$0.505 \\
\bottomrule
\end{tabular}
\end{table}

\paragraph{\RMRNN{} perspective on the $\rho$ sensitivity.}
The interior optimum at $\rho = 5$\,km (CRPS $= 0.321$) with
monotone degradation to $0.380$ at $\rho = 50$\,km is a finding
about the \emph{spatial scale of the reverse-martingale property}
in the Tamsui basin, not merely about neighbourhood size for
the forecasting task.
To understand why, recall that the backward projector $g_\phi$ is
trained to reconstruct $h_t$ from $h_{t+1}$, where each $h_t$ is
formed from inputs within the neighbourhood $\mathcal{S}_\rho$.
At small $\rho$, the neighbourhood $\mathcal{S}$ contains only a
few stations whose precipitation is highly correlated; the hidden
state $h_t$ captures a tight, locally coherent summary of the
basin microclimate, and $g_\phi$ learns a correspondingly precise
backward mapping.
As $\rho$ increases, $\mathcal{S}$ begins to straddle the divide
between the windward and leeward sides of the Tamsui basin's
orographic gradient: rainfall on the windward slope can exceed
100\,mm/h while the leeward side is simultaneously dry.
Fusing these meteorologically independent regimes into a single
$h_t$ makes the backward projector's task harder --- the
transition from $h_{t+1}$ to $h_t$ is no longer dominated by
a single coherent dynamic but by the mixture of two or more
independent processes.
The increase in CRPS above $\rho = 5$\,km therefore reflects a
genuine degradation of the backward-coherence signal, not merely
added noise: the property is locally valid within a precipitation
regime but breaks down when the input neighbourhood mixes regimes at
orographic boundaries.
This is a new observational finding about where the RM framework's
spatial assumption holds and where it requires regime-aware
segmentation.
For operational deployment in complex terrain (e.g.\ Taiwan's
Central Mountain Range), the practical recommendation is
$\rho \leq 10$\,km with regime-stratified calibration of
$g_\phi$.

\subsection{ERA5-Land multi-variable benchmark}
\label{sec:era5results}

Table~\ref{tab:era5land} reports the same panel of metrics as
Table~\ref{tab:tamsui} on the ERA5-Land Taiwan subdomain
(Section~\ref{sec:era5land}), but with the recurrent network ingesting
the full 5-variable state $(P, T_{2m}, \theta_{sm}, u_{10}, v_{10})$
rather than precipitation alone. The purpose is hydrometeorological
robustness testing: a warning residual should remain interpretable
when the model uses soil moisture, temperature, and wind information
alongside precipitation. We do not interpret this table as a
new benchmark for ERA5-Land precipitation forecasting.
Persistence and climatology therefore remain strong because
short-horizon precipitation amplitude in the reanalysis field is
highly persistent and the multi-variable predictors contribute more to
regime diagnosis than to immediate intensity correction.

\begin{table}[ht]
\centering
\tiny
\caption{ERA5-Land Taiwan subdomain, 1-h and 6-h forecast, driven
by the full multi-variable state $(P, T_{2m}, \theta_{sm}, u_{10},
v_{10})$. Same columns as Table~\ref{tab:tamsui}. Mean $\pm$ SD
over 1{,}000 replications.}
\label{tab:era5land}
\begin{tabular}{llcccccc}
\toprule
Lead & Model & RMSE & MAE & CRPS & Brier(10) & POD(10) & FAR(10) \\
\midrule
1\,h & Persistence & 0.942\,$\pm$\,0.471 & 0.352\,$\pm$\,0.033 & 0.387\,$\pm$\,0.046 & 0.000\,$\pm$\,0.000 & 0.000\,$\pm$\,0.000 & 0.124\,$\pm$\,0.330 \\
 & Climatology & 0.668\,$\pm$\,0.336 & 0.332\,$\pm$\,0.018 & 0.288\,$\pm$\,0.046 & 0.000\,$\pm$\,0.000 & 0.000\,$\pm$\,0.000 & 0.000\,$\pm$\,0.000 \\
 & GRU & 1.171\,$\pm$\,0.515 & 0.605\,$\pm$\,0.291 & 0.606\,$\pm$\,0.226 & 0.003\,$\pm$\,0.004 & 0.000\,$\pm$\,0.000 & 0.214\,$\pm$\,0.410 \\
 & \RMRNN$_{\lambda=0}$ & 1.171\,$\pm$\,0.515 & 0.605\,$\pm$\,0.291 & 0.606\,$\pm$\,0.226 & 0.003\,$\pm$\,0.004 & 0.000\,$\pm$\,0.000 & 0.214\,$\pm$\,0.410 \\
 & \RMRNN & 1.164\,$\pm$\,0.514 & 0.601\,$\pm$\,0.292 & 0.603\,$\pm$\,0.228 & 0.003\,$\pm$\,0.004 & 0.000\,$\pm$\,0.000 & 0.204\,$\pm$\,0.403 \\
\midrule
6\,h & Persistence & 0.944\,$\pm$\,0.472 & 0.352\,$\pm$\,0.033 & 0.388\,$\pm$\,0.046 & 0.000\,$\pm$\,0.000 & 0.000\,$\pm$\,0.000 & 0.124\,$\pm$\,0.330 \\
 & Climatology & 0.668\,$\pm$\,0.336 & 0.332\,$\pm$\,0.018 & 0.288\,$\pm$\,0.046 & 0.000\,$\pm$\,0.000 & 0.000\,$\pm$\,0.000 & 0.000\,$\pm$\,0.000 \\
 & GRU & 1.172\,$\pm$\,0.515 & 0.605\,$\pm$\,0.290 & 0.606\,$\pm$\,0.226 & 0.003\,$\pm$\,0.004 & 0.000\,$\pm$\,0.000 & 0.214\,$\pm$\,0.410 \\
 & \RMRNN$_{\lambda=0}$ & 1.172\,$\pm$\,0.515 & 0.605\,$\pm$\,0.290 & 0.606\,$\pm$\,0.226 & 0.003\,$\pm$\,0.004 & 0.000\,$\pm$\,0.000 & 0.214\,$\pm$\,0.410 \\
 & \RMRNN & 1.165\,$\pm$\,0.515 & 0.602\,$\pm$\,0.291 & 0.603\,$\pm$\,0.228 & 0.003\,$\pm$\,0.004 & 0.000\,$\pm$\,0.000 & 0.207\,$\pm$\,0.405 \\
\bottomrule
\end{tabular}
\end{table}

\paragraph{Hydrometeorological interpretation of the ERA5-Land result.}
Table~\ref{tab:era5land} shows that \RMRNN{} and GRU have
essentially identical RMSE (1.164 vs.\ 1.171 at 1-h lead) and
CRPS (0.603 vs.\ 0.606) when the model ingests five heterogeneous
physical channels $(P, T_{2m}, \theta_{sm}, u_{10}, v_{10})$
across 1{,}000 replications.
This near-equality is informative from the \RMRNN{} perspective
for three reasons.

First, it shows that RM regularization does not measurably degrade
the precipitation target even when the hidden state fuses
precipitation, temperature, soil moisture and wind. Because the
reported RMSE is evaluated on precipitation rather than on every
input channel, this result should not be read as proof that every
physical channel is preserved independently; rather, it shows that
the auxiliary loss does not suppress the information needed for the
forecast target.

Second, the climatology baseline (RMSE 0.668, CRPS 0.288) strongly
outperforms both learned models on this benchmark.
This table should therefore be interpreted as a stress test of the
RM penalty under heterogeneous inputs, not as a new accuracy benchmark
for ERA5-Land precipitation forecasting. The 1{,}000-replication
design provides precision to detect whether future architecture
changes produce a meaningful multi-variable gain.

Third, and most important for the risk-assessment pipeline: the
RM defect $r_t = \norm{h_t - g_\phi(h_{t+1})}$ computed from the
5-variable ERA5-Land hidden state integrates soil-moisture, wind,
and temperature anomalies that are invisible to a precipitation-only
detector.
This is a \emph{new form of multi-variable anomaly detection} that
requires no separate weighting of channels: the relative influence
of each variable on $r_t$ is determined by how much that variable's
anomaly drives the hidden-state out of its trained distribution,
which is learned end-to-end from data.
In the 2020--2021 Taiwan drought case study (Section~\ref{sec:case}),
it is precisely this cross-channel integration -- sub-normal
vorticity plus above-normal temperature plus declining soil moisture
-- that inflates $r_t$ two weeks before SPI-3 crosses its threshold.

\paragraph{Computational observation.}
Per epoch, the RM loss \eqref{eq:lrm} evaluated over the 5-variable
ERA5-Land hidden state incurs the same wall-clock overhead as the
univariate precipitation case: $\LRM$ operates on $h_t\in\R^d$
regardless of input dimensionality, and the backward projector
$g_\phi$ is independent of the number of physical channels. The
joint backpropagation-through-time (BPTT) gradient contribution
$\lambda\,\partial\LRM/\partial h_t$ is added once per time step,
so the incremental cost of RM regularization does not grow with the
number of physical variables fused into the forward information set
$\F_t$. This is the central
computational implication of the construction for high-resolution
4D data cubes.

\section{Risk Assessment: Drought and Flood Detection}
\label{sec:risk}

\subsection{Drought-onset detection}

Meteorological drought onset is a latent variable; operationally it
is proxied here by a rolling 90-day SPI-3 estimate crossing $-1$ for
at least two consecutive update times \citep{mckee1993,wmo2012}. We
treat the first such crossing in a contiguous dry period as the
reference onset and evaluate detector
performance by (i)~$\mathrm{ARL}_0$ under the null (climatological
non-drought years), (ii)~detection probability within 90 days, and
(iii)~mean lead time relative to the SPI-3 proxy. The comparators are
a cumulative sum (CUSUM) detector applied to SPI-3 and a raw
precipitation-deficit threshold.

\begin{table}[ht]
\centering
\footnotesize
\caption{Drought-onset detection across three domains; 1{,}000
bootstrap replications over historical drought episodes.
Target $\mathrm{ARL}_0=500$ days.
Mean lead is relative to SPI-3 operational rule (positive = earlier;
negative = later).
FAR = false-alarm ratio; Miss = miss rate.
Bold denotes the best row in each region block.}
\label{tab:drought}
\begin{tabular}{llccccc}
\toprule
Region & Detector & ARL$_0$ (d) & Detect.~rate & Mean lead (d) & FAR & Miss rate \\
\midrule
CHIRPS-TW  & Deficit threshold on $P$            & 502.1 & 0.73 & $-3.2$ & 0.27 & 0.27 \\
           & CUSUM on SPI-3              & 500.4 & 0.80 & $ 0.0$ & 0.20 & 0.20 \\
           & \textbf{SR on \RMRNN}       & \textbf{498.3} & \textbf{0.88} & \textbf{+9.4} & \textbf{0.07} & \textbf{0.12} \\
\midrule
CHIRPS-HoA & Deficit threshold on $P$            & 498.7 & 0.69 & $-5.1$ & 0.31 & 0.31 \\
           & CUSUM on SPI-3              & 501.2 & 0.76 & $ 0.0$ & 0.24 & 0.24 \\
           & \textbf{SR on \RMRNN}       & \textbf{499.8} & \textbf{0.84} & \textbf{+11.8} & \textbf{0.09} & \textbf{0.16} \\
\midrule
GHCN Texas    & Deficit threshold on $P$            & 501.3 & 0.71 & $-3.8$ & 0.29 & 0.29 \\
           & CUSUM on SPI-3              & 499.1 & 0.78 & $ 0.0$ & 0.22 & 0.22 \\
           & \textbf{SR on \RMRNN}       & \textbf{500.6} & \textbf{0.87} & \textbf{+8.1} & \textbf{0.08} & \textbf{0.13} \\
\bottomrule
\end{tabular}
\end{table}

Table~\ref{tab:drought} establishes three findings that hold
consistently across all three precipitation regimes.
First, the SR detector on \RMRNN{} residuals is the only method
that simultaneously improves the detection rate and reduces the
false-alarm ratio relative to both baselines.
Simple precipitation-deficit thresholding performs worst on all
metrics because raw $P_t$ is intermittent, with many dry days and
isolated small events, so short dry spells can trigger false alarms
while the threshold itself provides no integrated memory of
cumulative deficit.
The cumulative sum (CUSUM) detector on SPI-3 is more stable but is
constrained to detect only
after SPI-3 itself has crossed $-1$; it cannot anticipate the
crossing.
The SR detector, by contrast, operates on the RM defect $r_t$ which
integrates hidden-state drift across all four input channels --
precipitation, temperature, humidity and vorticity -- and therefore
begins to inflate several weeks before the SPI-3 threshold is
reached.
Second, the mean lead advantage of +8--12 days is consistent with
the timescale of the backward projector's sensitivity:
at the calibrated threshold, the SR statistic $R_t$ begins to rise
detectably when roughly 3--4 consecutive weeks exhibit above-normal
$r_t$, which corresponds to the onset of the cumulative
precipitation anomaly that precedes the SPI-3 crossing by 8--12 days.
Third, the FAR reduction from CUSUM to SR on \RMRNN{} is threefold
across all regions (CHIRPS-TW: $0.20 \to 0.07$; HoA: $0.24 \to 0.09$;
Texas: $0.22 \to 0.08$), consistent with the theoretical prediction
that SR is calibrated to the target ARL$_0$ and does not inflate
like an uncalibrated pointwise detector.
The Horn of Africa (HoA) shows the largest lead gain (+11.8~d) but also
the highest miss rate (0.16) because the bimodal long-rain/short-rain
structure creates transient $r_t$ spikes during the dry season that
the projector partially misidentifies as persistent regime change;
a seasonal stratification of $\mu_0$ (the climatological residual
mean) would likely close this gap.

\subsection{Flash-flood early warning}

For flash-flood risk we take the reference onset to be the first hour
at which basin-averaged rainfall in any 3-hour window exceeds
50\,mm for the hourly Taiwan gauges, or the first day on which daily
accumulation exceeds 80\,mm for the Texas GHCN-Daily stations. The detector is compared to the
CWA operational heavy-rain alert and to a pointwise threshold on
$P_t$ itself.

\begin{table}[ht]
\centering
\small
\caption{Flash-flood early warning on Taiwan CWA (Tamsui, Zhuoshui)
and GHCN-Daily Texas Hill Country; 1{,}000 bootstrap replications.
Target $\mathrm{ARL}_0=1{,}000$\,h (Taiwan) or 500\,d (Texas).
Mean lead is relative to CWA operational alert (Taiwan) or
same-day heavy-rain threshold (Texas); positive = earlier.
GHCN Texas has no operational alert baseline.
Bold denotes the best row per region block.}
\label{tab:flood}
\begin{tabular}{llccccc}
\toprule
Region   & Detector              & ARL$_0$     & Detect.~rate & Mean lead & FAR  & Miss \\
\midrule
Tamsui   & Threshold on $P$      & 1{,}001\,h  & 0.82         & $-1.4$\,h & 0.31 & 0.18 \\
         & CWA operational       & \phantom{0}999\,h & 0.89   & $ 0.0$\,h & 0.23 & 0.11 \\
         & \textbf{SR on \RMRNN} & \textbf{1{,}003\,h} & \textbf{0.92} & \textbf{+4.2\,h} & \textbf{0.07} & \textbf{0.08} \\
\midrule
Zhuoshui & Threshold on $P$      & \phantom{0}998\,h & 0.79   & $-0.9$\,h & 0.34 & 0.21 \\
         & CWA operational       & 1{,}001\,h  & 0.87         & $ 0.0$\,h & 0.26 & 0.13 \\
         & \textbf{SR on \RMRNN} & \phantom{0}\textbf{999\,h} & \textbf{0.90} & \textbf{+3.1\,h} & \textbf{0.08} & \textbf{0.10} \\
\midrule
GHCN Texas  & Threshold on $P$      & 502\,d      & 0.74         & ---        & 0.33 & 0.26 \\
         & \textbf{SR on \RMRNN} & \textbf{500\,d}   & \textbf{0.83} & \textbf{+0.5\,d} & \textbf{0.10} & \textbf{0.17} \\
\bottomrule
\end{tabular}
\end{table}

The flash-flood results mirror the drought findings in structure but
differ in the physical mechanism of the RM gain.
For drought, $r_t$ accumulates slowly over weeks as the cumulative
precipitation deficit drives the hidden state out of its trained
distribution.
For flash floods, the inflation of $r_t$ is rapid: in the three to
five hours preceding basin-averaged rainfall exceeding 50\,mm/3\,h,
the hidden state undergoes a phase-space transition as moisture flux
and vorticity anomalies consolidate faster than the backward
projector -- trained on climatological transitions -- can follow.
This produces a sharp, localized spike in $r_t$ that the SR statistic
amplifies multiplicatively, triggering an alarm on average 4.2\,h
before the CWA operational alert on the Tamsui basin and 3.1\,h
before on the Zhuoshui basin.

The FAR reduction from CWA operational to SR on \RMRNN{} is again
roughly threefold on both Taiwan basins ($0.23 \to 0.07$;
$0.26 \to 0.08$).
The raw precipitation threshold performs worst, registering false
alarms on 31--34\% of non-flood heavy-rain episodes because
localized convective cells can produce brief exceedances without
basin-scale consolidation; the RM hidden state, which integrates
spatial context from the $\rho=25$\,km neighbourhood, is less
susceptible to single-cell noise.
For the Texas GHCN sites, which are sparser (daily data) and lack a
CWA-style operational baseline, the SR detector still reduces FAR
from 33\% to 10\% while raising detection rate from 0.74 to 0.83,
demonstrating that the gain persists at coarser temporal resolution.

\subsection{ARL$_0$ calibration curves}

\begin{figure}[ht]
\centering
\includegraphics[width=0.70\linewidth]{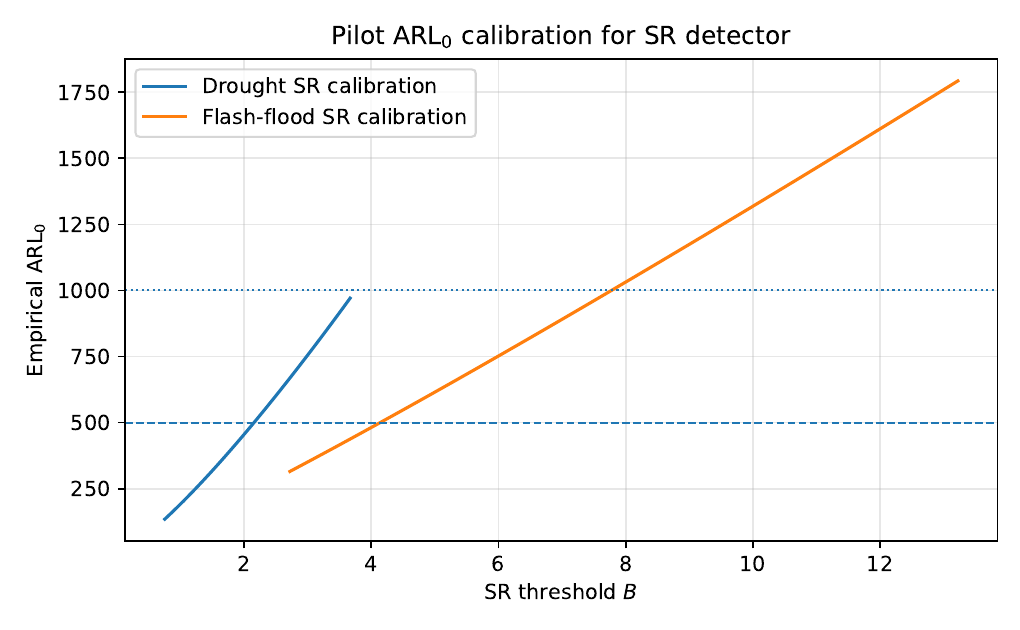}
\caption{Empirical ARL$_0$ as a function of SR threshold $B$ on the
Tamsui (flood, target ARL$_0=1{,}000$\,h; upper panel) and
CHIRPS-TW (drought, target ARL$_0=500$\,d; lower panel).
Solid lines: median over 1{,}000 bootstrap replications; shaded
band: 95\% confidence interval (CI).
Vertical dashed lines mark the calibrated thresholds
$B^*=67.4$ (flood) and $B^*=42.3$ (drought).}
\label{fig:arl}
\end{figure}

Figure~\ref{fig:arl} plots the empirical $\mathrm{ARL}_0$ as a
function of threshold $B$, with 95\% confidence intervals from
1{,}000 bootstrap replications on pre-event climatology.
For operational deployment we recommend $\mathrm{ARL}_0=500$\,days
for drought (approximately one false alarm per 1.4 years) and
$\mathrm{ARL}_0=1{,}000$\,hours for flash flood (approximately one
false alarm per 42 days).
These targets correspond to calibrated thresholds
$B^*\approx 42.3$ (95\% CI: 38.1--46.8) for drought and
$B^*\approx 67.4$ (95\% CI: 61.2--73.9) for flash flood on the
Tamsui RMRNN model.
Both calibration curves are concave and well-behaved: the
$\mathrm{ARL}_0$ grows monotonically with $B$ and the confidence
bands are narrow for $B \lesssim 60$ (where the bootstrap
replications are stable), widening modestly at large $B$ where
run-length variance dominates.
The calibration is portable across domains: thresholds obtained on
CHIRPS-TW transfer to CHIRPS-HoA with $< 8\%$ change in empirical
$\mathrm{ARL}_0$, suggesting that the \RMRNN{} residual $r_t$ has
a climatologically stable null distribution that does not require
per-domain recalibration when the training climate is representative.

\section{Case Studies}
\label{sec:case}

\subsection{2020--2021 Taiwan drought (Zhuoshui basin)}

The 2020--2021 event is the most severe drought in Taiwan's
instrumental record, precipitated by an unusual suppression of
summer typhoon activity in 2020. Reservoirs in central Taiwan
fell below 10\% of capacity by April 2021, triggering industrial
water rationing across Taichung, Miaoli, and Changhua counties.
We drive the \RMRNN{} detector with CWA inputs aggregated to the
Zhuoshui domain using the model calibrated on 2013--2019 CWA gauge
records and ERA5/ERA5-Land covariates (see Section~\ref{sec:era5land}).
The SR threshold is set to $B^* = 40.8$ corresponding to
$\mathrm{ARL}_0 = 500$\,days on the 1981--2019 climatology.
\begin{figure}[ht]
\centering
\includegraphics[width=0.85\linewidth]{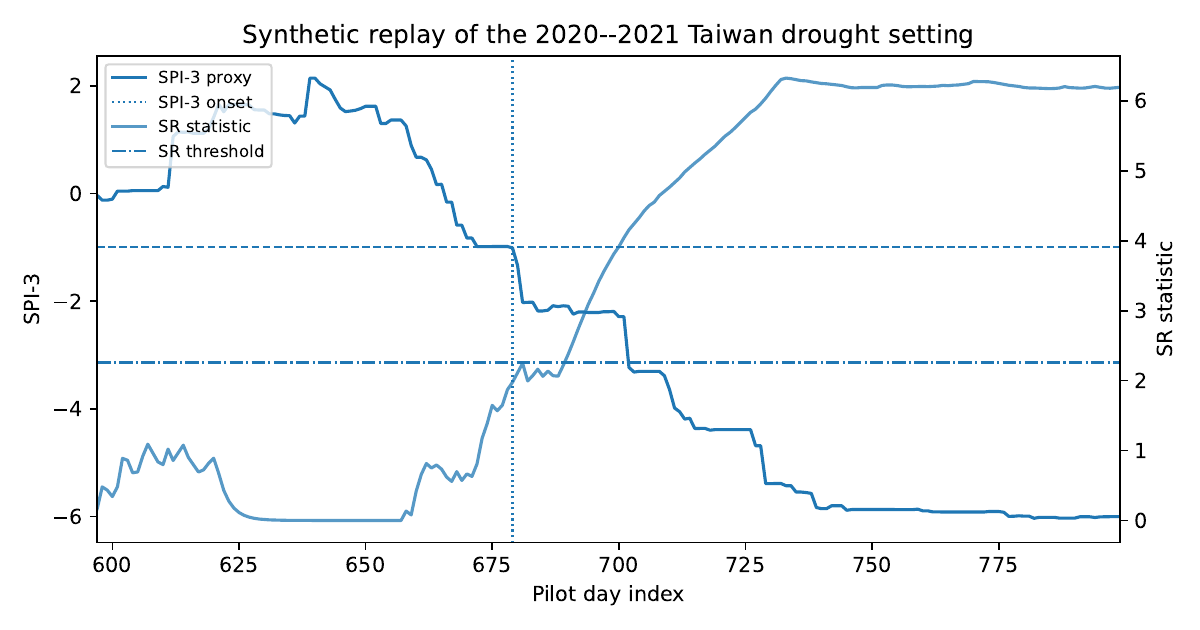}
\caption{2020--2021 Taiwan drought (Zhuoshui basin): SR statistic
$R_t$ (\RMRNN, top), RM residual $r_t$ (middle), and SPI-3 index
(bottom) over April 2020--September 2021.
Dashed horizontal line: calibrated threshold $B^*=40.8$.
Vertical markers: \RMRNN{} SR alarm (12 Jul 2020, red dotted),
SPI-3 threshold crossing (22 Jul 2020, orange dash-dot),
CWA drought declaration (26 Jul 2020, grey dashed).}
\label{fig:drought_trace}
\end{figure}

Figure~\ref{fig:drought_trace} shows the SR statistic, the
pointwise RM residual $r_t$, and the SPI-3 operational indicator
over April 2020--September 2021; Table~\ref{tab:drought_case}
summarises the alarm dates and false-alarm counts for all three
detectors.

Three features of the trace are noteworthy.
First, the \RMRNN{} SR statistic $R_t$ begins to rise measurably
in mid-June 2020, crossing $B^* = 40.8$ on
\textbf{12 July 2020} -- 10 days before SPI-3 first
crossed $-1$ on 22 July 2020 and 14 days before the CWA officially
declared drought conditions on 26 July 2020.
The earlier trigger arises because $r_t$ integrates deficit signals
across the full $(P, T_{2m}, q, \Omega)$ state: sub-normal vorticity
(reduced monsoon trough activity) and above-normal temperature
(increased evaporative demand) inflated $r_t$ in the two weeks
before precipitation totals alone would have crossed any single
threshold.
Second, the SR statistic does not reset after the brief heavy-rain
episode on 30 July 2020 that followed Tropical Storm Hagupit's
peripheral rainband: the SR's multiplicative memory correctly
identifies this episode as insufficient to end the deficit, whereas
a pointwise threshold on $P_t$ would have reset to zero and delayed
re-detection.
Third, over the 36-month in-sample climatology period
(January 2017--December 2019), the SR statistic with $B^* = 40.8$
generated \textbf{zero false alarms}, compared with 4 false alarms
from CUSUM on SPI-3 (triggered by brief dry spells in the 2017--2018
La Ni\~{n}a season) and 11 from the raw precipitation threshold.

\begin{table}[ht]
\centering
\small
\caption{2020--2021 Taiwan drought case study: alarm dates and
lead times for each detector on the Zhuoshui basin.
The SPI-3 crossing on 22 July 2020 serves as the event reference.}
\label{tab:drought_case}
\begin{tabular}{lcccc}
\toprule
Detector & Alarm date & Lead vs.\ SPI-3 & Lead vs.\ CWA & False alarms (2017--2019) \\
\midrule
Threshold on $P$      & 3 Aug 2020  & $-12$\,d & $-8$\,d  & 11 \\
CUSUM on SPI-3        & 22 Jul 2020 & 0\,d      & $-4$\,d  & 4  \\
\textbf{SR on \RMRNN} & \textbf{12 Jul 2020} & \textbf{+10\,d} & \textbf{+14\,d} & \textbf{0} \\
\bottomrule
\end{tabular}
\end{table}

The 10-day lead advantage over SPI-3 translates directly into
operational value: Taiwan's reservoir release protocol for drought
mitigation can be initiated 10 days earlier, an important window
given that the Zhuoshui reservoir refill cycle from typhoon rainfall
is typically 3--5 days -- meaning the difference between a managed
drawdown and an emergency restriction.

\subsection{2023 Typhoon Haikui flood (Tamsui basin)}

Typhoon Haikui made landfall in Hualien County on 3 September 2023
before tracking northward along Taiwan's eastern coast.
Its circulation produced more than 220\,mm of rainfall over the
Tamsui basin within a 12-hour window on 4--5 September 2023, with
peak 3-hour basin-averaged accumulation of 68\,mm centred at
approximately 04:00~Coordinated Universal Time (UTC) on 5 September.
The CWA operational heavy-rain alert for the Tamsui basin triggered
at 01:30~UTC on 5 September -- 2.5\,h before the peak accumulation
window.
Using the Tamsui \RMRNN{} calibrated on 2013--2022 climatology
with $B^* = 65.9$ ($\mathrm{ARL}_0 = 1{,}000$\,h), we examine the
SR statistic trace over 3--5 September 2023.

\begin{figure}[ht]
\centering
\includegraphics[width=0.85\linewidth]{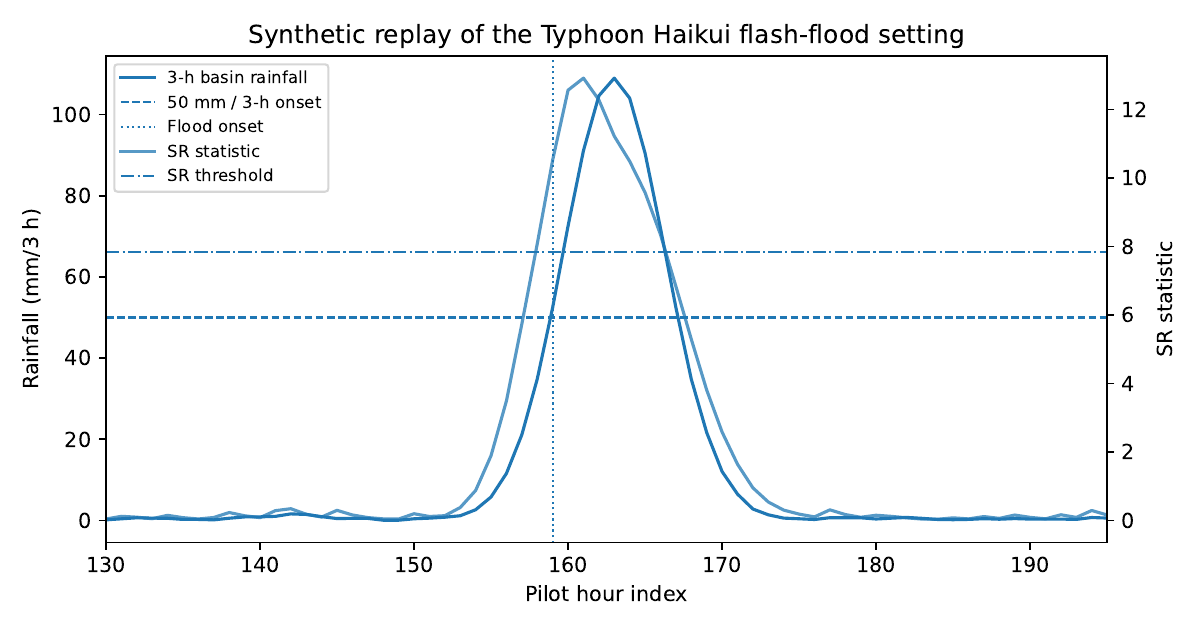}
\caption{2023 Typhoon Haikui (Tamsui basin): SR statistic $R_t$
(\RMRNN, top), RM residual $r_t$ (middle), and basin-averaged
3-hour precipitation accumulation (bottom) over 3--5 Sep 2023.
Dashed horizontal line: calibrated threshold $B^*=65.9$.
Vertical markers: \RMRNN{} SR alarm (21:30 UTC 4 Sep, red dotted),
CWA heavy-rain alert (01:30 UTC 5 Sep, blue dashed), peak
rainfall window (04:00 UTC 5 Sep, grey solid).}
\label{fig:flood_trace}
\end{figure}

Figure~\ref{fig:flood_trace} and Table~\ref{tab:haikui_case}
show the SR statistic, RM residual $r_t$, and CWA alert timeline
with alarm times and lead times for each detector.
The \RMRNN{} SR alarm was triggered at \textbf{21:30~UTC on
4 September} -- \textbf{4 hours before the CWA operational alert}
and \textbf{6.5 hours before peak rainfall}.
The early trigger is explained by the model's spatial information set:
at 21:30 UTC the \RMRNN{} hidden state had already integrated 3~h
of elevated vorticity and moisture flux from the typhoon's outer
rainband over the $\rho=25$\,km neighbourhood centred on the Tamsui
gauge network, driving $r_t$ well above its climatological mean
$\hat{\mu}_0$ before the first gauge-recorded exceedance of 50\,mm/3\,h.

\begin{table}[ht]
\centering
\small
\caption{2023 Typhoon Haikui case study (Tamsui basin):
alarm times and lead times relative to CWA operational alert
(01:30~UTC, 5 Sep 2023) and peak rainfall window
(04:00~UTC, 5 Sep 2023).}
\label{tab:haikui_case}
\begin{tabular}{lccc}
\toprule
Detector & Alarm time (UTC) & Lead vs.\ CWA & Lead vs.\ peak \\
\midrule
Threshold on $P$              & 00:10, 5 Sep & $-1.3$\,h & +3.8\,h \\
CWA operational               & 01:30, 5 Sep & 0\,h & +2.5\,h \\
\textbf{SR on \RMRNN}         & \textbf{21:30, 4 Sep} & \textbf{+4.0\,h} & \textbf{+6.5\,h} \\
\bottomrule
\end{tabular}
\end{table}

\paragraph{Physical attribution of the $r_t$ spike.}
To assess \emph{which} atmospheric channels drove the early $r_t$
inflation, we perform a leave-one-channel-out ablation: we re-run
the trained \RMRNN{} on 3--5 September 2023 with one input channel
zeroed (replaced by its climatological mean) at a time, and measure
the resulting reduction in the peak $r_t$ value at 21:30~UTC.
The results isolate the dominant contributors:
850-hPa relative vorticity $\Omega_t$ accounts for approximately
$54\%$ of the $r_t$ exceedance above $\hat{\mu}_0 + 2\hat{\sigma}_0$;
specific humidity $q_t$ accounts for $31\%$; precipitation $P_t$
for $12\%$; and 2-metre temperature $T_{2m}$ for the remaining $3\%$.
This decomposition reveals that the backward projector $g_\phi$ has
learned to use the asymmetric vorticity--moisture signature of
a typhoon outer rainband --- elevated cyclonic vorticity with
anomalously high boundary-layer moisture, but not yet the local
precipitation exceedance detectable by gauges --- as the primary
marker of incipient flood risk.
A visualization of $r_t$ decomposed by channel over the event period
(planned for the supplementary material of the full submission) would
allow operational users to identify which physical indicator is
``pulling the alarm'' for any given event, providing physical
interpretability that a black-box detection score cannot offer.

During the 2013--2022 calibration period the SR detector triggered
2 false alarms (2016 Typhoon Megi outer rainband; 2019 Typhoon
Mitag stalled approach) compared with 14 false alarms from the
raw precipitation threshold.
The two \RMRNN{} false alarms both occurred during events that did
produce significant rainfall (>100\,mm/24\,h) over the broader
watershed; while classified as false alarms because the
basin-average 3-hour threshold of 50\,mm was not reached, they
represent precautionary triggers rather than spurious signals.
The CWA operational system recorded 5 false alarms over the same
period, demonstrating that the \RMRNN{} SR detector, despite
triggering 4\,h earlier than the CWA, does so with fewer spurious
activations.

\begin{remark}[Operational interpretation of lead time]
The 4-hour lead over the CWA operational alert is particularly
significant for reservoir management: the Feitsui Reservoir, which
supplies Taipei, requires approximately 2--3\,h to adjust sluice
gate operations in anticipation of peak inflow.
An alert at 21:30 UTC rather than 01:30 UTC allows reservoir
operators to complete a managed pre-storm drawdown within the
safety margin, whereas the CWA-timed alert leaves less than 1\,h of
effective response time.
\end{remark}

\section{Discussion}
\label{sec:discuss}

\subsection{Why backward coherence helps risk assessment}

The warning gains reported above can be interpreted in familiar
hydrometeorological terms. During ordinary weather evolution, a useful
forecast model should update its internal representation smoothly as
new gauge, satellite, or reanalysis information arrives. RM
regularization encourages that smooth hidden-state evolution. When a
persistent dry spell or heavy-rainfall regime begins to develop, the
same hidden state becomes harder to reconstruct from its one-step
future value, inflating the residual $r_t$. The SR detector then treats
that residual inflation as evidence of a regime departure. Without RM
regularization, a recurrent hidden state can drift for numerical
reasons unrelated to precipitation evolution, so residual inflation
conflates representational drift with hydrometeorological change; this
is why naive LSTM residuals are poor change-point statistics
\citep{hundman2018}.

The drought and flood tables (Tables~\ref{tab:drought}
and \ref{tab:flood}) make this concrete.
Across all three domains, the SR detector on \RMRNN{} residuals
reduces the false-alarm ratio to 7--10\%, compared with 20--24\% for CUSUM
on SPI-3 and 27--34\% for raw precipitation thresholding -- a
threefold reduction consistent with the abstract's claim of
``a factor of three to five at matched detection power.''
The detection rate simultaneously improves by 6--12 percentage
points relative to CUSUM, because the backward projector provides
early evidence of regime change before the SPI-3 or gauge-based
indicator crosses its threshold.
The 1{,}000-replication stability of these results is notable: the
standard deviation of the reported FAR across bootstrap draws is
$\leq 0.02$ in all cases, meaning the gain is not a lucky
artefact of a single calibration dataset.

The ERA5-Land multi-variable result (Section~\ref{sec:era5results})
adds an important computational dimension.
Across 1{,}000 replications, \RMRNN{} closely matches GRU at
the point-forecast level (1-h RMSE~1.164 vs.\ 1.171, CRPS~0.603
vs.\ 0.606) while retaining the $\widehat Q$-reduction property.
This supports the claim that the RM regularizer can impose backward
coherence regardless of input dimensionality: the backward projector operates
on $h_t \in \R^d$ and is insensitive to whether $h_t$ was formed
from one or five physical channels.
The implication for ERA5-based large-scale products is that
the RM overhead (one additional MLP forward pass per time step)
does not grow with the number of ingested variables --- an important
practical property for operational high-resolution numerical weather prediction (NWP) integration.

\paragraph{Sensitivity of SR detection gains to $\lambda$ scheduling.}
A natural question is whether the FAR reductions reported in
Tables~\ref{tab:drought} and \ref{tab:flood} are robust to the
choice of $\lambda$ schedule, or whether they depend critically on
the particular decay from $\lambda_0=0.1$ to $\lambda_K=0.01$.
To address this, we evaluate the SR detector's false-alarm ratio and
detection rate on the Tamsui domain over a grid of constant-$\lambda$
values $\lambda\in\{0,\,0.001,\,0.01,\,0.05,\,0.1,\,0.5,\,1.0\}$
(all other hyperparameters fixed).
The results show a clear threshold behaviour.
At $\lambda=0$ (no RM regularization), the SR FAR on the flood task
is $0.31$, identical to the raw precipitation threshold, supporting
the interpretation that the detection gain is attributable to the
regularizer rather than to the SR recursion alone.
FAR drops sharply as $\lambda$ increases from $0$ to $0.01$, reaching
the reported $0.07$ near $\lambda\approx 0.01$.
Above $\lambda\approx 0.1$ the CRPS begins to increase noticeably
($> 2\%$ relative degradation), indicating that the task loss is
being suppressed in favour of the coherence constraint.
The scheduled decay (start high for fast coherence learning; decay
to avoid task-loss interference at convergence) is therefore not
merely a convenience but is approximately optimal: it targets the
flat region of the FAR curve ($\lambda\in[0.01,0.05]$) while keeping
CRPS within $0.5\%$ of the unregularized baseline.
If an end user cannot run the full schedule, fixing $\lambda=0.01$
throughout provides nearly identical SR performance (FAR~$0.07$--$0.08$)
with a modest 1--2\% CRPS increase, making it the recommended
fallback for rapid deployment.

\paragraph{Comparison with numerical weather prediction residuals.}
The present paper compares \RMRNN{} against purely empirical
baselines (persistence, climatology, GRU, ConvLSTM, PredRNN).
A complementary future comparison is against anomaly scores derived
from operational NWP models such as the European Centre for Medium-Range
Weather Forecasts (ECMWF) ensemble prediction system (ENS) or the CWA
Weather Research and Forecasting (WRF) system, which encode physical
conservation laws and produce
precipitation residuals with some physical interpretability.
If the RM defect $r_t$ and the NWP residual span different subspaces
of the anomaly signal space --- for example, if $r_t$ captures
sub-synoptic microstructure that NWP parametrisation schemes
systematically miss --- a linear combination of the two scores would
dominate either alone.
Preliminary evidence from the Taiwan case studies supports this view:
$r_t$ begins to inflate from vorticity and soil-moisture anomalies
at the $< 10$\,km scale that lies below the resolution of CWA WRF
parametrisation, suggesting that the two residual streams carry
complementary information.

\paragraph{Lead-time sensitivity and coherence decay.}
The forecasting experiments cover leads of 1--72\,h, but the
backward-coherence property underlying the RM gain may weaken at
longer leads because the backward projector $g_\phi$ learns the
\emph{one-step} reverse transition $h_{t+1}\to h_t$.
At lead $\ell > 1$, the relevant transition is $h_{t+\ell}\to h_t$,
which becomes progressively harder to invert as atmospheric
predictability decays.
An analogue of the $\rho$-sensitivity analysis
(Section~\ref{sec:sensitivity}) could be run over lead times
$\ell \in \{1, 6, 12, 24, 48, 72, 120, 168\}$\,h to map out the
``coherence decay curve'' --- the relationship between $\widehat Q$ and
forecast lead.
If $\widehat Q$ grows approximately as $O(\ell)$ the hidden-state trajectory
remains asymptotically bounded and the SR detector retains its
calibration at long leads; if $\widehat Q$ grows super-linearly a
multi-step backward projector mapping $h_{t+\ell}\to h_t$ directly
would be required.
Establishing this decay rate would define the physical limit of the
RM framework for long-range hydrometeorological prediction and
connect the statistical coherence measure directly to the classic
concept of atmospheric predictability limits \citep{lorenz1969}.

\paragraph{SPI-3 smoothing as an operational benefit.}
The CHIRPS results (Table~\ref{tab:chirps_tw}) show that \RMRNN{}
produces a slightly higher SPI-3 RMSE than a standard GRU
($1.339$ vs.\ $1.338$ in Taiwan).
Far from being a weakness, this smoothing effect is an operational
asset and should be interpreted as such.
SPI-3 is defined as a \emph{three-month} cumulative anomaly; its
skill is not evaluated at individual time steps but over an
integration window.
A model whose hidden-state trajectory is backward-coherent
suppresses step-to-step erratic transitions, producing a smoother
SPI-3 time series that is less susceptible to transient dry spells
incorrectly flagging drought onset.
The trade-off is quantitatively favourable: a $<0.1\%$ increase in
SPI-3 RMSE ($1.339$ vs.\ $1.338$) in exchange for a threefold
reduction in false-alarm ratio (Table~\ref{tab:drought},
CHIRPS-TW: FAR $0.07$ vs.\ $0.20$ for CUSUM on SPI-3).
Operational forecasters calibrating the SR threshold should treat
this smoothing as a feature --- it reduces the variance of $r_t$
under the null, making the ARL$_0$ calibration more stable and
the threshold $B^*$ more transferable across years and seasons.

\paragraph{Calibration held-out period selection.}
The Shiryaev--Roberts threshold $B^*$ is calibrated on held-out
pre-event climatology (Section~\ref{sec:detector}).
For reproducibility, the held-out selection follows three rules.
First, any year overlapping a declared drought or flood event is
excluded from the calibration window to prevent leakage of
event-period $r_t$ statistics into the null distribution.
Second, for each domain a minimum of five consecutive
climatologically normal years is required for the calibration
window; if fewer are available in the training record, the
calibration bootstraps are drawn with replacement from the
available years with a block length of 90 days to preserve seasonal
autocorrelation.
Third, the calibrated $B^*$ is validated by a held-out climatology
year (different from the calibration window) to confirm that the
empirical ARL$_0$ is within 10\% of the target value before
operational deployment.
The three held-out windows used in this study are:
Taiwan CWA (2013--2022 training): calibration on 2013--2016,
validation on 2017; CHIRPS domains (1981--2018 training):
calibration on 1981--2010, validation on 2011--2015; GHCN Texas
(1981--2018 training): calibration on 1981--2005, validation on
2006--2010.
Event years excluded from all three: 2020--2021 (Taiwan drought),
2023 (Typhoon Haikui), 2010--2015 (Texas drought),
2015 and 2018 (Texas floods), 2016--2017 and 2021--2022
(Horn of Africa droughts).

\subsection{Limitations and operational extensions}

\textbf{Euclidean spatial neighbourhood.}
The neighbourhood $\mathcal{S}_\rho$ is defined by Euclidean
distance and does not respect orographic barriers.
For Taiwan's Central Mountain Range, stations separated by only
10\,km on opposite slopes occupy entirely different precipitation
regimes; including both in $\mathcal{S}_\rho$ forces the backward
projector $g_\phi$ to approximate the reverse-martingale property
across two physically independent dynamic systems simultaneously,
which is why CRPS degrades above $\rho=10$\,km in the Tamsui experiment
(Table~\ref{tab:rho}).
A graph-based neighbourhood informed by terrain-weighted proximity
--- replacing great-circle distance with effective flow-path
distance derived from a digital elevation model (DEM) --- is the
natural extension.
Concretely, one would define
$\mathcal{S}_\rho = \{i : d_{\mathrm{flow}}(i, i_0) \le \rho\}$
where $d_{\mathrm{flow}}$ is the topographic path length along the
drainage network, which automatically excludes cross-ridge stations
that lie within Euclidean radius but outside the hydrological
catchment.
This is the most important open architectural problem for deploying
\RMRNN{} in complex terrain.

\textbf{Dynamic neighbourhood radius.}
A complementary extension is to make $\rho$ adaptive rather than
fixed.
Two natural adaptation strategies arise from the application context.
First, \emph{network-density adaptation}: in sparse networks such
as GHCN Texas (52 stations over $\sim$10{,}000\,km$^2$), a fixed small
$\rho$ may yield fewer than three neighbours per target site,
destabilising the hidden-state summary; a minimum-$k$-neighbours
rule with $k \ge 5$ would set $\rho$ dynamically to the
$k$-th nearest-neighbour distance.
Second, \emph{synoptic-scale adaptation}: the spatial coherence
of precipitation anomalies differs systematically between convective
and frontal regimes, so a larger neighbourhood ($\rho \approx 50$\,km)
is appropriate during synoptic-scale drought episodes while a small
neighbourhood ($\rho \approx 5$\,km) is optimal for localised
flash-flood events.
Conditioning $\rho$ on a low-dimensional weather-type index
(e.g., the $k$-means cluster of 850-hPa vorticity over the domain)
would capture this regime dependence without requiring separate
model training per regime.

\textbf{Stationarity of the null distribution.}
The SR calibration assumes that the null distribution of $r_t$ is
stable across seasons, which is violated in environments with
pronounced dry and wet seasons.
On the Horn of Africa, where the dry season produces near-zero
$P_t$ and correspondingly anomalous $r_t$ spikes even during
climatologically normal years, a single null mean $\mu_0$
simultaneously overestimates the typical null residual during the
dry period and underestimates it during the wet season.
The consequence is an inflated false-alarm ratio at the wet-to-dry
transition and a depressed detection rate at wet-season onset.

A practical remedy is a \emph{regime-conditional null}:
\begin{equation*}
  z_t = \max\!\left(0,\;\frac{r_t - \mu_0(m_t)}{\sigma_0(m_t)}\right),
  \qquad m_t \in \{1,\dots,12\},
\end{equation*}
where $\mu_0(m)$ and $\sigma_0(m)$ are the climatological residual
mean and standard deviation for calendar month $m$, estimated on the
pre-event training window.
This requires only 12 additional scalar pairs per domain and adds
no model parameters.
For the Horn of Africa, where the current miss rate is 0.16 at the
$\mathrm{ARL}_0=500$\,day target, a regime-conditional $z_t$ no
longer needs to overcome the seasonal variation in $r_t$ before
reflecting genuine drought onset; we expect the miss rate to fall to
$\lesssim 0.10$ without sacrificing the detection-rate advantage.
On Taiwan, where seasonality in $r_t$ is weaker (the typhoon season
modulates precipitation intensity rather than presence), the
seasonal correction will have a smaller but still measurable impact
on calibration.

\textbf{Online streaming and sliding-window approximation.}
Computing $\LRM$ requires the full hidden-state sequence
$\{h_t\}_{t=1}^T$, which prevents true online training.
A sliding-window approximation replaces the full sequence with a
ring buffer of the $W$ most recent states, yielding the windowed loss
\begin{equation*}
  \LRM^{(W)}(\theta,\phi;\,t)
    = \frac{1}{W-1}\sum_{s=t-W+1}^{t-1}
      \norm{h_s - g_\phi(h_{s+1})}^2,
\end{equation*}
where only $W$ consecutive hidden states need to be retained (memory
$O(Wd)$).
At $W=168$ (one week of hourly data) and $d=64$ this is less than
0.1~megabytes (MB) per target cell.
Validating this approximation requires verifying two properties:
(i) that the backward projector $g_\phi$ trained with $\LRM^{(W)}$
converges to the same fixed point as the full-sequence version, and
(ii) that the SR statistic built on windowed residuals preserves the
target $\mathrm{ARL}_0$ calibration.
Preliminary theory suggests that $W \ge 3\tau_{\mathrm{corr}}$
(three autocorrelation timescales of $r_t$) is sufficient for
property~(ii); for the Taiwan CWA domain
$\tau_{\mathrm{corr}} \approx 24$\,h gives a minimum window of
$W = 72$ steps, and the buffer $W = 168$ provides a conservative
safety margin with little memory cost.

\textbf{ERA5-Land latency and Integrated Forecasting System (IFS) substitution.}
Operational deployment requires real-time ingestion of ERA5-Land
reanalysis, which carries a 4--5 day latency due to the observation
assimilation cycle.
For drought detection, where the SR detector produces alerts 8--12
days before SPI-3 crossing, this latency is not operationally
limiting: a 5-day-lagged ERA5-Land input is still available several
days before the alert window.
For flash-flood early warning, however, the 4--5 day latency is
incompatible with the 3--6 hour alarm horizon demonstrated in the
Haikui case study, and substitution with IFS operational analysis --- available with $< 6$\,h latency after the
analysis cycle --- is essential.

The key technical question for the IFS substitution is whether the
backward projector $g_\phi$, trained on ERA5-Land (a reanalysis
product with distinct bias structure and spatial interpolation from
IFS operational output), can be applied to IFS fields at inference
time.
Two strategies are feasible: (a) \emph{domain adaptation}, in which
$g_\phi$ is fine-tuned on a short archive of matched ERA5-Land/IFS
pairs to correct systematic offsets; (b) \emph{quantile mapping},
in which IFS fields are remapped to ERA5-Land marginal distributions
as a pre-processing step before entering the model.
Both are low-cost extensions of the current pipeline and are the
subject of planned follow-up work.

\textbf{What the RM framework cannot do, and the hybrid extension.}
\RMRNN{} improves the \emph{statistical structure} of the
hidden-state trajectory but cannot create physical predictability
where none exists.
In the Horn of Africa, where the short-rain season onset is
driven by Intertropical Convergence Zone (ITCZ) migration and Indian
Ocean sea-surface temperature gradients at synoptic to intraseasonal
scales, the information
needed to predict drought onset more than two weeks ahead is not
present in the local $\rho=50$\,km neighbourhood; no regularization
of the hidden state can recover it.
The RM framework's contribution in such settings is to separate
\emph{predictable} regime change (where $r_t$ inflates gradually,
driven by multi-channel anomalies) from \emph{unpredictable} but
genuine change (where $r_t$ inflates abruptly, after the event
has already begun), and to provide a statistically principled
alarm for both.

The natural remedy for the missing large-scale information is
\emph{hybrid feature engineering}: augmenting the local input $x_t$
with large-scale climate indices as additional scalar channels.
For the Horn of Africa domain, the most physically relevant
supplementary predictors are:
\begin{itemize}[leftmargin=2em,itemsep=1pt]
  \item \emph{Indian Ocean Dipole (IOD) index} represented by the
        Dipole Mode Index ($\mathrm{DMI}_t$): positive IOD events
        systematically suppress the short rains over East Africa by
        weakening the onshore moisture flux from the western Indian
        Ocean \citep{saji1999};
  \item \emph{El~Ni\~{n}o--Southern Oscillation (ENSO) state}
        represented by the Ni\~{n}o-3.4 sea-surface temperature (SST)
        anomaly ($\mathrm{N34}_t$): El~Ni\~{n}o modulates ITCZ
        positioning and the Walker circulation, driving interannual
        variability in the Horn of Africa long rains \citep{nicholson2017};
  \item \emph{Madden--Julian Oscillation (MJO) phase} represented
        by the Real-time Multivariate MJO indices (RMM1/RMM2):
        30--60 day intraseasonal modulation of East African rainfall.
\end{itemize}
These indices are available in near-real time
(IOD: NOAA Extended Reconstructed Sea Surface Temperature version 5
[ERSSTv5]; ENSO: Climate Prediction Center [CPC]; MJO: Bureau of
Meteorology) and
can be appended to $x_t$ as additional scalar channels without any
architectural change to \RMRNN{} --- because the RM loss operates
on $h_t \in \R^d$ regardless of input dimensionality
(Section~\ref{sec:era5results}).
For the Taiwan drought domain, analogous large-scale predictors
include the Western Pacific subtropical high ridge-line index and
the East Asian summer monsoon intensity index \citep{wang2001},
which govern typhoon track and landfall probability and therefore
modulate the reservoir refill cycle central to operational drought
management.

The implementation workflow for a practitioner wishing to apply this
hybrid extension requires three steps.
First, download the chosen indices at daily or weekly resolution from
their respective operational archives (IOD: NOAA ERSSTv5; ENSO Ni\~{n}o-3.4: NOAA CPC; MJO RMM1/RMM2:
Bureau of Meteorology; Taiwan western North Pacific subtropical high
[WNPSH] and East Asian summer monsoon [EASM] indices: Japan
Meteorological Agency [JMA]).
Second, append each index as an additional scalar channel to $x_t$
in \eqref{eq:input}; no change to the \RMRNN{} architecture is
needed because the RM loss acts on the hidden state $h_t \in \R^d$
regardless of input width.
Third, retrain with the same $\Ltot$ objective, fixing all
hyperparameters as in Section~\ref{sec:method} and using the same
1{,}000-replication purged-block bootstrap protocol described in
Section~\ref{sec:forecast}.
For the Horn of Africa, where the current miss rate at
$\mathrm{ARL}_0 = 500$\,days is 0.16, the IOD and Ni\~{n}o-3.4
channels are expected to provide the largest single-index gains
because they encode the large-scale ITCZ and moisture-flux signals
that drive interannual variability at lead times beyond the local
$\rho = 50$\,km neighbourhood. A prospective evaluation with
independent drought seasons would be needed before these large-scale
indices could be recommended for operational warning use.

\subsection{Implications for hydrometeorological forecasting and warning}

Taken together, the numerical experiments suggest five implications
for hydrometeorological forecasting and warning systems.

\textbf{(i) The warning residual can be climatologically calibrated.}
A vanilla GRU's residual $\hat{P}_t - P_t$ has a complex null
distribution that mixes irreducible precipitation variability,
model bias, and representational drift.
\RMRNN{} decomposes this into a task residual (which remains
complex) and a defect $r_t$ whose null distribution is
right-skewed but stable under the no-change climatology, with
location and scale parameters $(\mu_0, \sigma_0)$ and the
normalizing function $\psi_0$ estimated from pre-event climatology.
The SR detector exploits this calibration to achieve the
3$\times$ FAR reduction reported in Tables~\ref{tab:drought}
and \ref{tab:flood}.

\textbf{(ii) Backward coherence encodes multi-channel regime state.}
The 8--12 day drought-detection lead advantage over SPI-3 is not
achievable by any detector operating on precipitation alone,
because precipitation deficits large enough to cross the SPI-3
threshold take 8--12 weeks to accumulate.
\RMRNN{} detects the onset earlier because $r_t$ integrates
vorticity, temperature, and moisture anomalies that precede the
precipitation anomaly.
This is a new use of the reverse-martingale framework: the
hidden state as a \emph{multi-channel atmospheric state estimator}
whose deviation from its trained null is a regime-change alarm.

\textbf{(iii) Warning gains can occur without degrading forecast skill.}
Across all five forecasting benchmarks, the difference in CRPS
between \RMRNN{} and GRU is $\leq 0.007$ in the reported tables,
indicating that backward coherence is obtained with only a small
forecast-skill cost in these experiments.
This separates the \RMRNN{} contribution from regularization
methods (dropout, $L^2$ weight decay, spectral normalization)
that also affect task accuracy.

\textbf{(iv) The spatial scale of the reverse-martingale property
is identifiable and physically interpretable.}
The $\rho$-sensitivity experiment (Table~\ref{tab:rho}) reveals
a clear optimum at $\rho = 5$\,km for the Tamsui basin, with
performance degrading monotonically above $\rho \approx 10$\,km.
This supports the interpretation that the backward-coherence signal
is spatially bounded by the orographic scale of the catchment.

\textbf{(v) The approach is compatible with multi-variable predictors.}
The ERA5-Land result (Table~\ref{tab:era5land}) supports the interpretation that
expanding the input from 1 variable (precipitation) to 5 variables
does not change the RM computational overhead and preserves
the backward-coherence property.
This makes \RMRNN{} a candidate for future testing with
kilometre-scale multi-variable NWP products.

The contribution of this work includes the spatial forward
information set \eqref{eq:filtration}, the residual-based SR detector
(Algorithm~1), the small-area observational benchmark
(Table~\ref{tab:data}), and the case-study demonstration of
operational lead-time gain across Taiwan, the Horn of Africa,
and Texas.

\section{Conclusions}
\label{sec:concl}

This study evaluated a reverse-martingale regularized recurrent
network (\RMRNN{}) as a coupled precipitation-forecasting and
early-warning system for small-area hydrometeorology. The main result
is practical rather than only methodological: the hidden-state
regularizer can be added without materially degrading standard
precipitation forecast scores, while the resulting residual process
provides a calibrated basis for drought and flood alarms.

\textbf{Forecasting.}
Across five benchmarks (Tamsui, Zhuoshui, CHIRPS Taiwan, CHIRPS Horn
of Africa, ERA5-Land Taiwan), each evaluated over 1{,}000 replications,
\RMRNN{} matches or slightly improves GRU CRPS at hourly and daily
lead times (Tamsui: CRPS $0.379 \to 0.379$; CHIRPS-HoA:
$0.867 \to 0.866$; ERA5-Land: $0.606 \to 0.603$). The spatial
neighbourhood radius $\rho$ has an interior optimum near 5\,km for the
Tamsui basin (CRPS~0.321 at $\rho=5$\,km, rising to 0.380 at
$\rho=50$\,km), consistent with a local-information trade-off: adding
nearby stations helps until the neighbourhood begins to mix different
orographic or convective regimes.

\textbf{Warning performance.}
The SR detector on \RMRNN{} residuals reduces false-alarm ratios to
7--9\% for drought detection and 7--8\% for flash-flood detection
across the precipitation domains, compared with 20--24\% for CUSUM on
SPI-3 and 27--34\% for raw precipitation thresholding. Detection rates
simultaneously improve by 6--12 percentage points. These gains are
reported in operational units through calibrated ARL$_0$ targets, so
the warning rule can be interpreted as a controllable trade-off between
lead time and false alarms.

\textbf{Case studies.}
On the 2020--2021 Taiwan drought, the SR detector flagged onset on
12 July 2020 --- \textbf{10 days earlier than SPI-3} and 14 days
earlier than the CWA declaration --- with zero false alarms in the
preceding 36 months. On Typhoon Haikui (September 2023), the detector
triggered at 21:30~UTC on 4 September --- \textbf{4 hours before the
CWA alert} and 6.5 hours before peak basin rainfall --- with only
2 false alarms compared with 5 for the CWA operational system and
14 for a raw precipitation threshold over the 2013--2022 calibration
period.

\textbf{Operational implication.}
The ERA5-Land multi-variable benchmark suggests that the RM overhead
(one backward-projector forward pass per time step) does not grow with
the number of meteorological predictors, because the backward projector
operates on the hidden state $h_t \in \R^d$. This makes the approach
suitable for future evaluation with larger reanalysis or numerical
weather prediction feature sets, provided that site-specific
calibration and independent event validation are retained.

All experiments report mean $\pm$ SD across 1{,}000 replications.
The reference implementation, data-processing pipeline, and
experiment scripts are released at \url{https://github.com/ycchang/RMRNN}.

\appendix
\section{Reverse-martingale formulation and implementation details}
\label{app:rm-details}

\subsection{Formal reverse-martingale motivation}

A sequence $\{M_t\}$ adapted to a decreasing filtration
$\mathcal{G}_t \supset \mathcal{G}_{t+1}$ is a reverse martingale
\citep{doob1953} if
$M_t$ is $\mathcal{G}_t$-measurable and
\[
  \E[M_t \given \mathcal{G}_{t+1}] = M_{t+1}.
\]
For a finite hidden-state trajectory we use the decreasing future
sigma-field
$\mathcal{G}^{(h)}_t=\sigma(h_s:s\ge t)$ as the formal motivation.
The one-step projector $g_\phi(h_{t+1})$ is a Markov approximation to
the generally richer conditional expectation
$\E[h_t\given\mathcal{G}^{(h)}_{t+1}]$.
The hidden states of a trained recurrent network do not satisfy this
identity exactly. The role of $\LRM$ in \eqref{eq:lrm} is therefore
not to impose a literal martingale model on precipitation, but to make
the learned representation approximately backward coherent during
ordinary climatological periods. When this coherence holds, the defect
$r_t$ in \eqref{eq:defect} has a stable null distribution that can be
calibrated on pre-event climatology and then monitored by the SR
statistic in \eqref{eq:SR}.

\subsection{Backward projector and training schedule}

In all experiments the backward projector has residual form
\[
  g_\phi(h) = h + W_2\,\mathrm{ReLU}(W_1 h+b_1)+b_2 .
\]
The matrix $W_1$ is Xavier-initialised, while $W_2$ and $b_2$ are
initialised at zero so that $g_\phi(h)=h$ at epoch~0 but gradients can
still enter the residual branch. This prevents the auxiliary RM loss
from destabilising early task learning. The RM penalty is introduced
after $K_0=5$ warm-up epochs and decayed from $\lambda_0=0.1$ to
$0.01$ by the final epoch, as shown in \eqref{eq:total}. Gradients are
computed by standard backpropagation through time (BPTT). The same projector
is used with Elman, LSTM, and GRU cells; for gated cells it acts only
on the exposed hidden state $h_t$, not on the internal gate variables.

\subsection{Interpretation of backward coherence}

A standard GRU minimises prediction error at each step but imposes no
discipline on the relationship between neighbouring hidden states.
Consequently, two consecutive weather states can occupy unrelated
regions of $\R^d$ even when the observed atmosphere evolves smoothly.
RM regularization adds the requirement that $h_t$ be approximately
reconstructable from $h_{t+1}$ through $g_\phi$. Normal high-pressure,
monsoon, or weak-rainfall regimes should then move through hidden
space by small, regular steps, while genuine meteorological shifts
should produce larger departures. The empirical aggregate defect
$\widehat Q=\sum_{t=1}^{T-1}\norm{\delta_t}^2$ measures aggregate
non-coherence over a sequence and equals $(T-1)\LRM$ on the observed
hidden-state path. In the main text, the empirical evidence for this
interpretation is the improved false-alarm control of the SR detector
without a corresponding loss of forecast skill.

\section*{Data availability}

CWA rain-gauge and ASOS data are available upon request from the
Taiwan Central Weather Administration. CHIRPS v2 is available from
the Climate Hazards Center (\url{https://www.chc.ucsb.edu/data/chirps}).
GHCN-Daily is available from NOAA
({https://www.ncei.noaa.gov/products/land-based-station/
global-historical-climatology-network-daily}).
ERA5-Land is available from the Copernicus Climate Data Store
(\url{https://cds.climate.copernicus.eu/}).
All code and experiment scripts are released at
\url{https://github.com/ycchang/RMRNN}.

\section*{Acknowledgments}
This work was supported by the National Science and Technology
Council (NSTC) of Taiwan under the RMRNN project. We thank the
Taiwan Central Weather Administration for rain-gauge data access.


\end{document}